\documentclass{aa}  

\usepackage{graphicx}
\usepackage{txfonts}
\usepackage{color}
 
\newcommand{\brac}[1]{\left\langle #1 \right\rangle}

\newcommand{\lH}{\ell_\mathrm{H}}
\newcommand{\bnab}{\bm{\nabla}}
\newcommand{\dd}{\mathrm{d}}
\usepackage{bm}
\newcommand{\rev}[1]{#1}

\begin{document} 
 
   \title{Self-organisation in protoplanetary disks}
   \subtitle{global, non-stratified Hall-MHD simulations}

   \author{William B\'ethune\inst{1,2}
          \and
          Geoffroy Lesur\inst{1,2} 
          \and
          Jonathan Ferreira\inst{1,2}}

   \institute{Univ. Grenoble Alpes, IPAG, F-38000 Grenoble, France
   			\and CNRS, IPAG, F-38000 Grenoble\\
   			\email{william.bethune@ujf-grenoble.fr}
             }

   \date{Received date; accepted date}

  \abstract
   {Recent observations revealed organised structures in protoplanetary disks, such as axisymmetric rings or horseshoe concentrations evocative of large-scale vortices. These structures are often interpreted as the result of planet-disc interactions. However, these disks are also known to be unstable to the magneto-rotational instability (MRI) which is believed to be one of the dominant angular momentum transport mechanism in these objects. It is therefore natural to ask if the MRI itself could produce these structures without invoking planets. }
   {The nonlinear evolution of the MRI is strongly affected by the low ionisation fraction in protoplanetary disks
   The Hall effect in particular, which is dominant in dense and weakly ionised parts of these objects, has been shown to spontaneously drive self-organising flows in shearing box simulations. Here, we investigate the behaviour of global MRI-unstable disc models dominated by the Hall effect and characterise their dynamics. }
   {We validate our implementation of the Hall effect into the PLUTO code with
predictions from a spectral method in cylindrical geometry. We then perform 3D
unstratified Hall-MHD simulations of keplerian disks for a broad range of Hall,
ohmic and ambipolar Elsasser numbers. }
   {We confirm the transition from a turbulent to an organised
state as the intensity of the Hall effect is increased. We
observe the formation of zonal flows, their number depending on the available magnetic
flux and on the intensity of the Hall effect. For intermediate Hall intensity, the flow
self-organises into long-lived magnetised vortices. Neither the addition of a
toroidal field nor ohmic or ambipolar diffusion drastically change this picture in the range
of parameters we have explored. }
   {Self-organisation by the Hall effect is a robust phenomenon in global non-stratified simulations. It is able to quench
turbulent transport and spontaneously produce axisymmetric rings or sustained vortices. The ability of these structures to trap dust particles in this configuration is demonstrated. We conclude that Hall-MRI driven organisation is a plausible scenario which could explain some of the structures found in recent observations. }

   \keywords{accretion, accretion disks --
                magnetohydrodynamics (MHD) --
                protoplanetary disks --
                stars: formation --
                turbulence
               }

   \maketitle
%

\section{Introduction}

Protoplanetary discs are the birth place of planets. There are now
many observations available, from millimetre wavelengths \citep{ALMA15} to the 
infrared \citep{BJ15}. Surprisingly, these observations indicate that
protoplanetary discs have complex internal structures, with features like
spiral arms \citep{M12,BJ15}, asymmetric traps \citep{MD13} and rings
\citep{ALMA15}.

The origin of these structures is largely unknown. Most of the theoretical
models developed to date involve at least one planet carving a gap
\citep{MD13,DP15} or exciting spiral density waves \citep{BJ15}. For this
reason, structures are usually seen as a consequence of planet formation.
However, the presence of Jupiter mass planets in very young discs such as HL 
tau is questionable since it implies a very fast formation scenario. It is therefore
natural to ask if these structures could appear without planets, and even
be one of the driving mechanism behind planet formation. To answer this question,
one has to focus on the dynamics of a planet-free disc, a long-standing
problem in theoretical astrophysics.

Accretion discs are believed to be subject to the magnetorotational
instability (MRI), a magnetohydrodynamic (MHD) instability appearing spontaneously in 
magnetised discs in Keplerian rotation \citep{BH91}. The relevance of this instability
in the context of protoplanetary disc is however debated \citep{TFG14}. In particular,
these objects are so cold that they cannot maintain a large ionisation fraction \citep{G96}. 
This has led to the concept of magnetically ``dead'' zones where the ionisation fraction
of the disc is so low that the standard MHD approach breaks down and the MRI potentially vanishes.

A large effort has recently been devoted to the study of these weakly ionised regions, typically
located at radii larger than 1 astronomical unit (a.u.) which are precisely the regions we are now capable
of resolving observationally. It has been shown that in this regime,
three ``non-ideal'' effects have to be taken into account in the MHD framework \citep{BT01,KB04}:
Ohmic diffusion, Ambipolar diffusion and the Hall drift. It is well known that both Ohmic and Ambipolar
diffusion acts to damp and potentially eliminate the MRI by decoupling the flow from magnetic field
lines \citep{BB94,J96,PT97}, hence the historical denomination of dead zones. On the other hand, 
the Hall effect leads to new branches of instabilities \citep{W99,BT01,K08} due to its dispersive nature. 

The saturation of the MRI and the outcome of a disc subject to these non ideal effects is still the subject
of intense debate. Stratified models including Ohmic and ambipolar diffusion only have shown that the disc
could become essentially laminar (i.e. not turbulent) in the midplane, with a strong magnetically-driven jet at the disc surface \citep{BS13,SX13,GTNN15}. Adding Hall effect leads to a significantly different picture for radii\footnote{These 
radii have been computed assuming a minimum mass solar nebula \citep[MMSN,][]{H81} 
for the disc density and temperature structure. Real discs are likely to deviate 
significantly from this model, so these radii should be taken with care when 
comparing to specific astrophysical objects.} between 1 and 10 a.u. (where Hall effect
dominates over diffuse processes) a large-scale 
midplane stress is found, with a strong field polarity dependence 
and no turbulence \citep{LKF14,B14} whereas the outer disc is found to be essentially insensitive to 
the Hall effect \citep{B15} though some polarity dependence could persist \citep{SLK15}. 

The results of \cite{LKF14} and \cite{B14}, and in particular the presence of a large-scale
``laminar'' stress is both an important and potentially problematic result. In effect, 
these simulations were done in the
stratified shearing box framework \citep{HGB95}, with a very limited radial extension 
(typically a few scale height). So \cite{LKF14} and \cite{B14} results indicate that the MRI is indeed
active in these simulations, but it tends to produce structures \emph{larger} than the 
simulation box. As a matter of fact, unstratified shearing box models have already indicated 
that Hall-dominated MRI is prone to produce large-scale structures such as zonal fields and flows
\citep[][hereafter KL13]{KL13} in wide enough boxes. All of these elements tend to point toward the fact that the Hall effect
could be an efficient mechanism to trigger self-organisation in protoplanetary discs.

In this paper, we explore this hypothesis using full 3D cylindrical models of accretion discs
dominated by the Hall-effect. Our aim is not to reproduce the exact structure of a disc with its detailed
ionisation equilibrium but to test the hypothesis of self-organisation in a global model, free of the artefacts found in shearing-box simulations. We start by describing in details the physical framework
and some remarkable facts concerning Hall dynamics. We then depict our numerical model and present the
tests performed to validate it along with a reproduction and discussion of the multifluid models of \cite{KD14}. Next, we move to the core of the paper with the presentation and characterisation of several Hall-dominated models, putting emphasis on the physical mechanisms driving self-organisation. 
We then explore some potential treats to these mechanisms such as diffusive effects and more complex field configurations. We conclude by discussing the link between these results and recent observations along with extensions of this work to more realistic numerical setups.


\section{Framework} \label{}

\subsection{Physical model} \label{sec:physmod}

We want to model a thin disk of partially ionised gas orbiting a central mass
and threaded by a weak magnetic field. At the scales considered, our problem
fits in the non-ideal magnetohydrodynamic (MHD) framework. In this attempt to
study the Hall effect in global disk simulations, we wish to concentrate on the
midplane dynamics, thereby neglecting vertical stratification effects. Using
cylindrical coordinates $(r,\varphi,z)$, we take the gravitational force to be
oriented radially and assume periodicity in the vertical direction on a
scale $h$. 
Radial stratification is eliminated by using an isothermal disk model with a
flat density distribution, and by taking the initial magnetic field to be
constant and uniformly vertical over the whole domain. 

The relevant equations for our model are those of inviscid, compressible
non-ideal MHD. We denote by $\rho$ the bulk density, $\bm{v}$ the bulk
velocity, $P = c_s^2\rho$ the pressure with an isothermal sound speed $c_s$,
and $\bm{J} = \nabla \times \bm{B}$ the electric current deduced from the
magnetic field $\bm{B} = \mid B \mid \bm{e_b}$. Gravity is oriented along the
cylindrical radius as $\bm{g}=-(1/r^2) \bm{e_r}$. The three relevant non-ideal
MHD effects, namely Ohmic dissipation, the Hall effect and ambipolar diffusion,
are characterized by their diffusivities $\eta_{\mathrm{O}}$,
$\eta_{\mathrm{H}}$ and $\eta_{\mathrm{A}}$ respectively \citep[e.g.][]{W07}.
With these definitions, the dynamical equations read:
\begin{align}
&\partial_t \rho = - \bm{\nabla \cdot} \left[ \rho \bm{v} \right],
\label{eqn:dyn-rho}\\
&\partial_t \left[ \rho \bm{v} \right] = - \bm{\nabla \cdot} \left[ \rho \bm{v
\otimes v} \right] - \bnab P + \bm{J \times B} + \rho\bm{g},
\label{eqn:dyn-v}\\
&\partial_t \bm{B} = \bm{\nabla \times} \left[ \bm{v \times B} -
\eta_{\mathrm{O}} \bm{J} - \eta_{\mathrm{H}} \bm{J \times e_b} +
\eta_{\mathrm{A}} \bm{J \times e_b \times e_b} \right].
\label{eqn:dyn-b}
\end{align}

In order to characterize the intensity of the Hall induction term, we use the Alfv\'en velocity $\bm{v_A} = \bm{B} / \sqrt{\rho}$ and introduce the dimensionless parameter 
\begin{equation}
\mathcal{L} \equiv \lH / h
\end{equation}
ratio of the Hall length $\lH\equiv\eta_H/v_A$ as defined in \cite{KL13}
(hereafter KL13) and the geometric scale height of the disk. The Hall length
is very convenient as it only depends on the microphysical properties of
the plasma and not on the field strength. In a plasma made of electrons, singly
charged ions and neutrals, the Hall length is given by
\begin{align}
\ell_H&=\Bigg(\frac{c^2m_i}{4\pi
e^2n_e}\Bigg)^{1/2}\Bigg(\frac{\rho}{\rho_i}\Bigg)^{1/2}\\
&=0.14\,\Bigg(\frac{\rho}{10^{-9}\,\mathrm{g.cm}^{-3}}\Bigg)^{-1/2}\Bigg(\frac{\xi}{10^{-12}}\Bigg)^{-1}\,\mathrm{a.u},
\end{align}
where $m_i$ is the ion mass, $n_e$ is the electron number density and $\rho_i$
is the ion density. The numerical value has been obtained assuming a plasma made of electrons, heavy ions and neutral molecules (mean molecular mass $\mu=2.34\,m_\mathrm{H}$) with an ionisation fraction $\xi$. The precise calculation of the Hall length involves solving a
complex chemical network and depends on the presence and size distribution of
grains, which is beyond the scope of this paper. We note however that typical
values for $\mathcal{L}$ range from $\mathcal{L}\sim100$ at 5 a.u. down to
$\mathcal{L}\sim 10^{-2}$ at 100 a.u. for $1\,\mu\mathrm{m}$ size grains
\citep[e.g.][]{SLK15}, smaller grains leading to larger values of
$\mathcal{L}$. The Hall effect being a dispersive term, it induces a strong
Courant-Friedrichs-Lewy (CFL) constraint on the timestep of our explicit
scheme. For this reason, we restrict ourselves to simulations with
$\mathcal{L}\le 10$ to keep computation times within reasonable bounds.

Most of our simulations neglect ohmic and amibipolar diffusion to focus on
Hall-MHD only. We however briefly explore the impact of these diffusion terms
on our results in section \ref{sec:threats}. It is customary to quantify the importance of
diffusion terms via the Lundquist and Elsasser numbers respectively defined by
\begin{equation}
\mathcal{S}_{\mathrm{O},\mathrm{A}} \equiv \frac{v_A h}{\eta_{\mathrm{O},\mathrm{A}}}, \qquad
\Lambda_{\mathrm{O},\mathrm{A}} \equiv \frac{v_A^2}{\eta_{\mathrm{O},\mathrm{A}} \Omega}, 
\end{equation}
where the index stands for ohmic or ambipolar diffusion. When setting the value of these parameters, it will be with respect to the initial Alfv\'en velocity $B_0/\sqrt{\rho_0}$.

In this paper, we wish to exclude most stratification effects. Since we do not solve for the chemistry of the disk, fixing a constant $\lH$ gives a natural extension of the local simulations of KL13. We choose to set $\eta_O$ constant so that $\mathcal{S}_{\mathrm{O}}$ describes a disk that is either globally stable or unstable to the MRI \citep{J96}. Most previous studies on ambipolar diffusion were conducted at a given $\Lambda_{\mathrm{A}}$, but with a constant Alfv\'en velocity this corresponds to a disk gradually stabilised by the diffusivity $\eta_{\mathrm{A}} \propto r^{3/2}$. We choose instead to use a constant ion-neutral coupling time $\tau_\mathrm{A}$ so that $\eta_\mathrm{A}(\bm{r},t)=\tau_{\mathrm{A}}v_A(\bm{r},t)^2$ everywhere and at all time. With an initially constant Alfv\'en velocity, the initial $\mathcal{S}_{\mathrm{A},0} = h/\tau_{\mathrm{A}} v_{A,0}$ is constant too, making the disk globally unstable and facilitating comparison with the ohmic case.

\subsection{Simplified model of Hall dynamics} \label{sec:simo}

We point out some properties derived from the dynamical equation
\eqref{eqn:dyn-b}, and refer the reader to the section 4 of KL13 for a discussion about the following model. 

Discarding the ambipolar and ohmic terms, and assuming a constant Hall length, the induction
equation of Hall MHD reads
\begin{align}
\nonumber \partial_t \bm{B} &= \bm{\nabla \times} \left[  \bm{v\times B}-\lH
\bm{J \times B}\right] \\
 \label{eqn:induc_easy}& =\bm{\nabla \times} \left[  \bm{v\times B}\right]- \lH
\bm{\nabla \times} \left[ \bm{\nabla \cdot} \left(\bm{B \otimes B} \right) \right]. 
\end{align}
For illustrative purposes we use the shearing box model to remove curvature terms; we can then average over the periodic vertical and azimuthal directions and obtain the simplified equation
\begin{align}
\begin{split}
\partial_t \brac{\bm{B}_z}_{yz} &=  \brac{\bm{\nabla \times} \left[\bm{v \times
B}\right]_z} - \lH \brac{\bm{\nabla \times} \left[ \bm{\nabla \cdot} \left(\bm{B
\otimes B} \right) \right]_z} \\
&\simeq \eta_t \frac{\partial^2}{\partial x^2} \brac{B_z} - \lH
\frac{\partial}{\partial x^2} \brac{B_xB_y} \label{eq:inucinde}
\end{split}
\end{align}
where we have approximated the ideal MHD term by a turbulent resistivity
\citep{LL09}. This expression shows that in the presence of the Hall effect, the
horizontal Maxwell stress appears explicitly in the induction equation. Since
this component of the stress also drives the transport of angular momentum
\citep{BP99}, this expression implies a tight connection between the transport
of angular momentum and that of vertical magnetic flux tubes. These relations
will prove useful in understanding our results.

\subsection{Linear stability} \label{sec:linstab}

We recall that in Hall MHD, differentially rotating flows embedded in a
magnetic field can be subject to the Hall-shear instability (HSI), different in
essence from the magneto-rotational instability \citep{K08}. Indeed,
angular-momentum conservation is central to the MRI mechanism, whereas the HSI
relies only on electric currents and needs not disturb the bulk of the flow. One
other fundamental difference is the dependence of the HSI on the orientation of
the magnetic field with respect to the rotation axis. In the case favorable to
instability ($\bm{\Omega\cdot B}>0$), it can be shown \citep{SLK15} that the
HSI develops when the local shear frequency is larger than the whistler
frequency at the considered wavenumber $k$, that is :
\begin{equation}
\label{eq:HSIcondition}
q \Omega  \geq \lH v_A k^2
\end{equation}
with $q \equiv \vert\, \dd \log \Omega / \dd \log r \,\vert = 3/2$ in our non-stratified keplerian case. Given a constant $\lH$, we see that the flow will be Hall-shear stable for all wave numbers down to the minimal $k_0 = 2\pi / h$ for Alfv\'en velocities above the critical value
\begin{equation}
\label{eq:HSIeasy}
v_{A,\mathrm{crit}} \equiv \frac{3 \Omega_0 h^2 }{8 \pi^2 \lH}\Bigg(\frac{r}{r_0}\Bigg)^{-3/2}. 
\end{equation}

\subsection{Diagnostics} \label{}

We use brackets to denote space averaging, with subscript indicating the
coordinate over which a function is integrated: 
\begin{align}
\brac{f}_{z} &\equiv \frac{1}{h} \int_{z = -h/2}^{h/2} f \, \dd z, \\
\brac{f}_{r\varphi} &\equiv \frac{2}{(r_{out}^2-r_{in}^2)\Delta\varphi} \int_{r
=
r_{in}}^{r_{out}}\int_{\varphi = 0}^{\Delta \varphi} f \, r \, \dd\varphi\, \dd
r, 
\end{align}
with $r_{in}$ and $r_{out}$ delimiting the active computational domain and
$\Delta \varphi$ the angular extent of the disk. Subscripts will be omitted when obvious
from the context. We use an overline for the time averaging of an already
volume-averaged quantity :
\begin{equation}
\overline{f} \equiv \frac{1}{T} \int_{t=t_0}^{t_0+T} \brac{f} \, dt.
\end{equation}
As a characterization of the spatial coherence of the flow, we use the
auto-correlation function 
\begin{align}
\mathcal{A}_{y,z}[f](x) \equiv \frac{\int f(t,y,z) f(t+x,y,z) \, \dd t}{\int
f(t,y,z)^2\,\dd t}
\end{align}
and define a normalized correlation factor
\begin{equation}
\mathcal{C}_z[f](r,\varphi) \equiv \frac{1}{h} \int_{z = -h/2}^{h/2} 
\mathcal{A}_{r,\varphi}[f](z)\, \dd z
\end{equation}
equal to unity where the correlation length is equal to the height $h$. We also
use $\mathcal{C}_{\varphi}[f](r,z)$ defined by substituting $\varphi$ and $z$ and
replacing $h$ by the angular extent $\Delta \varphi$. 

Following \cite{BP99}, the turbulent Reynolds stress tensor is defined with the
velocity fluctuations about the density weighted averaged flow: $\mathcal{R}
\equiv \rho \tilde{\bm{v}} \otimes \tilde{\bm{v}}$ where $\tilde{\bm{v}} =
\bm{v} - \brac{\rho \bm{v}}_{\varphi z} / \brac{\rho}_{\varphi z}$. The Maxwell stress tensor is simply defined as $\mathcal{M} \equiv -
\bm{B} \otimes \bm{B}$. We introduce two dimensionless measures of turbulent
stress; the first one is the usual $\alpha$ parameter of \cite{SS73}:
\begin{equation}
\label{eq:defalphaSS}
\alpha_\mathrm{SS} \equiv \frac{\mathcal{R}_{r\varphi} +
\mathcal{M}_{r\varphi}}{\rho c_s^2}. 
\end{equation}
Since we work in a cylindrical setup, an alternative definition for $\alpha$ is
possible using $\Omega$ and the characteristic length $h$. We thus define 
\begin{equation}
\label{eq:defalpha}
\alpha_{\mathcal{R}} \equiv \frac{\mathcal{R}_{r\varphi}}{\rho\Omega^2 h^2 },
\qquad  \alpha_{\mathcal{M}} \equiv \frac{\mathcal{M}_{r\varphi}}{\rho\Omega^2
h^2 }, \qquad \alpha \equiv \alpha_{\mathcal{R}} + \alpha_{\mathcal{M}}.
\end{equation}
It should be emphasized again that in a disk in true hydrostatic equilibrium,
$c_s^2=\Omega^2 h^2$ so that $\alpha_\mathrm{SS}=\alpha$. In a cylindrical
system such as ours, this equality does not hold anymore and one has to pick
\eqref{eq:defalphaSS} or \eqref{eq:defalpha}. In the following, we use
\eqref{eq:defalpha} as our main diagnostic since the fluctuations are
essentially incompressible, making the sound speed less relevant physically
than the geometrical velocity $\Omega h$.

\subsection{Units}

We take the inner radius $r_0$ and the keplerian velocity at the inner radius
$v_0$ as distance and velocity units, resulting in a natural frequency unit
$\Omega_0 \equiv v_0/r_0$. Time is expressed relative to the orbital period at
the inner radius $T_0 \equiv 2\pi / \Omega_0$, and density relative to the
initial constant density $\rho_0$. Magnetic field strength is expressed as an
Alfvén velocity $B / \sqrt{\rho_0}$, and the initial vertical magnetic field is always denoted $B_0$. 
We will also use the initial thermal pressure $P_0 \equiv \rho_0 c_s^2$.


\section{Method and validation} \label{sec:compmeth}

\subsection{Description} \label{sec:compdesc}

\subsubsection{Numerical scheme} \label{sec:numschem}

The dynamical equations \eqref{eqn:dyn-rho}-\eqref{eqn:dyn-b} are explicitly
integrated in time with a modified version of the finite volume code PLUTO
\citep{M07} in 3D cylindrical geometry. The magnetic field is evolved with the
constrained transport method \citep{EH88}, preserving $\nabla \cdot \bm{B} = 0$
to machine precision. 

Our implementation of the Hall effect is the same as that described in Appendix
A of \cite{LKF14}: the Hall induction flux is incorporated in
a conservative manner
within a modified HLL Riemann solver. Since the Hall effect introduces
dispersive waves, the whistler speed required when computing Godunov fluxes is
truncated at the grid scale in the direction under consideration. The
face-centered electric currents must be provided to this solver as an external
parameter. These are computed by finite difference of either the
volume-averaged or the face-centered magnetic field of neighbouring cells,
according to the cylindrical expression for the curl operator. 

We use a second-order accurate Runge-Kutta time-integration scheme, a piecewise
linear space reconstruction of the fields within each cell, and the Van Leer
slope limiter. The FARGO module \citep{MFSK12} is employed to substantially
increase the initial timesteps and reduce the numerical diffusivity due to the
advection by the mean Keplerian flow.

\subsubsection{Grid and boundary conditions} \label{sec:gridcond}

Our computational domain is cylindrical, truncated in the azimuthal direction to
quarter disks $\Delta \varphi = \pi/2$ for all runs except one $2\pi$ full disk, saving computational time while revealing global effects
already. The ratio of the cylinder height $h$ to its inner radius $r_0$ is
fixed to $1/4$ for all simulations, consistently with our thin disk assumption.
We set the outer radius at $5 r_0$, giving an aspect ratio of $16$. The sound
speed is fixed to $10\%$ of the keplerian velocity at the inner radius, so that
our geometrical scale height $h$ is of the order of the hydrostatic pressure scale height
$c_s / \Omega$ in the computational domain, being equal at $r\simeq 1.8r_0$.
This domain is meshed with 32 grid cells in the vertical direction, 512 uniformly distributed cells in the radial and
azimuthal directions (2048 azimuthal cells for the $2\pi$ run), giving a square mesh in the $\left(r,z \right)$ plane and
an almost cubic grid at $r \approx 2.5 r_0$. 

All runs are initialised with a density $\rho = \rho_0$ on the whole domain, a
keplerian azimuthal velocity field $v_{\varphi} = r^{-1/2}$ and uniform $B_z =
B_0$. The vertical and azimuthal boundary conditions are
periodic. Special care was taken with the radial boundary conditions, for they
can drastically affect the physics in the computational domain. The same type
of conditions are imposed at the two radial boundaries; unless otherwise stated
they are as follows: the radial and vertical velocities vanish as well as the
radial derivative of the density; the azimuthal velocity is set to its initial
keplerian value; the azimuthal component of the magnetic field is forced to zero
while the vertical component is set to $B_0$, and the radial component follows
from the divergence free condition. 

Boundary conditions alone allow the development of large-scale quasi-steady
spiral waves within the domain in purely hydrodynamical setup. These
undesirable effects appeared for a variety of boundary conditions, so we chose
to add damping buffer zones to smooth the transition from the active domain to
the boundaries and smooth these structures out. The buffers span a radial
extent $h$ from both radial boundaries, where all hydrodynamical fields are
linearly relaxed over time. They are excluded from all subsequent
space-averagings. 

In the half of the buffer closest to the boundary, all velocity components are
brought back to their initial value with an exponential relaxation: $\bm{v} \mapsto
\bm{v} - S_1(r) (\bm{v}-\bm{v_0})\delta t/\tau$, with characteristic time scale
$\tau = 0.125 / \Omega(r)$, and with a linear modulation $S_1(r)$ such that the
relaxation is fully active at the boundary and vanishes at the middle of the
buffer. 

In the whole buffer, the density is relaxed in the following manner. First, the average radial density distribution $\brac{\rho}_{\varphi z}(r)$ is computed. Let $r_b$ be the radius of the edge of the current buffer. In each cell the
density is relaxed exponentially in time to the arithmetic mean between the
local average density and the average at the edge of the buffer: 
\begin{equation}
\rho(r,\varphi,z) \mapsto \rho(r,\varphi,z) - S_2(r)
\left[\brac{\rho}(r)-\brac{\rho}(r_b)\right] \frac{\delta t}{2\tau},
\end{equation}
where $S_2(r)$ is again a linear modulation equal to one at the boundary and
zero at $r_b$. 

This separation into two sub-buffers helps preventing an accumulation of mass
between the buffer and the active domain, avoiding the formation of a local
minimum of potential vorticity and keeping the flow Rossby stable
\citep{LCN99}. Also, relaxing the density to its average value from the active
domain mimics outflow conditions and allows long time behavior such as a
density drop by accretion to take place. 

We impose a constant resistivity $\eta_b=4\times 10^{-4}$ in the buffers,
stabilizing the MRI for typical magnetic field intensities without affecting
the CFL condition. Moreover, in Hall MHD simulations we linearly decrease the
Hall length from the active domain and make it vanish at the boundaries,
avoiding unnecessary Hall drift waves at the interface. 

Finally, after a first round of simulations we found that magnetic flux losses
were sometimes too large to maintain a near statistically steady system. This prevented us from computing proper time averages and  comparing our results with previous works. Part of this monotonic decrease in net flux comes from turbulent
accretion / excretion of mass out of the computational domain, but it is amplified by the Hall term
\begin{align}
\partial_t \langle B_z\rangle_{r\varphi}\simeq\oint \lH
\dd\bm{l\cdot}(\bm{\nabla\cdot \mathcal{M}}),
\end{align}
always negative since we damp the Maxwell stress in the buffers with ohmic
diffusion. \rev{This term inputs negative flux from the boundaries, which spreads to the active domain and progressively stabilizes the flow. }
We thus renormalise the total vertical magnetic flux at each time
step in the active domain  $B_z(r,\varphi,z) \mapsto B_z(r,\varphi,z) -
\brac{B_z} + \brac{B_0}$. This way, we obtain quasi-steady states in all of our regimes. 
\rev{In the ideal-MHD case, the amount of magnetic flux thus added varies between $10\%$ and $30\%$ of the initial magnetic flux over $200T_0$, the larger flux losses corresponding to the higher transport rates, occurring for higher initial net flux. In the strong-Hall regime, less than $5\%$ of the total flux is lost from $50T_0$ to $200T_0$ in the organized phase. These flux losses are small on a dynamical time scale, and we verified that the maximal turbulent stress attained in time were the same with and without this procedure. The qualitative level of self-organisation was also consistent between runs with and without the flux renormalisation. }
One caveat of this method is that mass losses should enhance the
turbulent stress in time as the relative importance of magnetic to thermal
pressure increases \citep{HGB95}.

\subsection{Validation}  \label{sec:compvalid} \label{sec:specmeth}

We developed a Chebyshev pseudo-spectral method to compute the MRI eigenmodes
in axisymmetric cylindrical coordinates, described in appendice
\ref{app:spectral}. It includes viscosity and all three non-ideal MHD effects,
and was tested in ideal, ohmic, Hall and ohmic plus Hall configurations. 

The method returns the axisymmetric eigenmodes of the non-ideal MHD equations,
linearized about a stationnary keplerian flow with uniform vertical magnetic
field. In this section, we take the computational domain to be the unit square
$\left(r,z\right) \in \left[1,2\right]^2$. The boundary conditions are periodic
in $z$, with vanishing $v_r$, $v_{\varphi}$ and $B_{\varphi}$, $\partial_r v_z$
and $\partial_r B_z$ at the radial boundaries. 

We compare the profiles predicted with a resolution of 256 spectral modes to
the result of direct numerical simulations (DNS) on a $256^2$ grid. The initial
and boundary conditions implemented in PLUTO are the same as those of the
spectral method, with a white noise of amplitude $\delta v / c_s < 10^{-10}$
seeded in the $v_r$ and $v_z$ fields. The fastest growing MRI mode rapidly
dominates, and the growth is slow enough that we can extract the radial
profiles at a given height, normalize them so that they have the same amplitude
and compare them to linear predictions. 

\begin{figure}[h]
\centering
\includegraphics[width=\hsize]{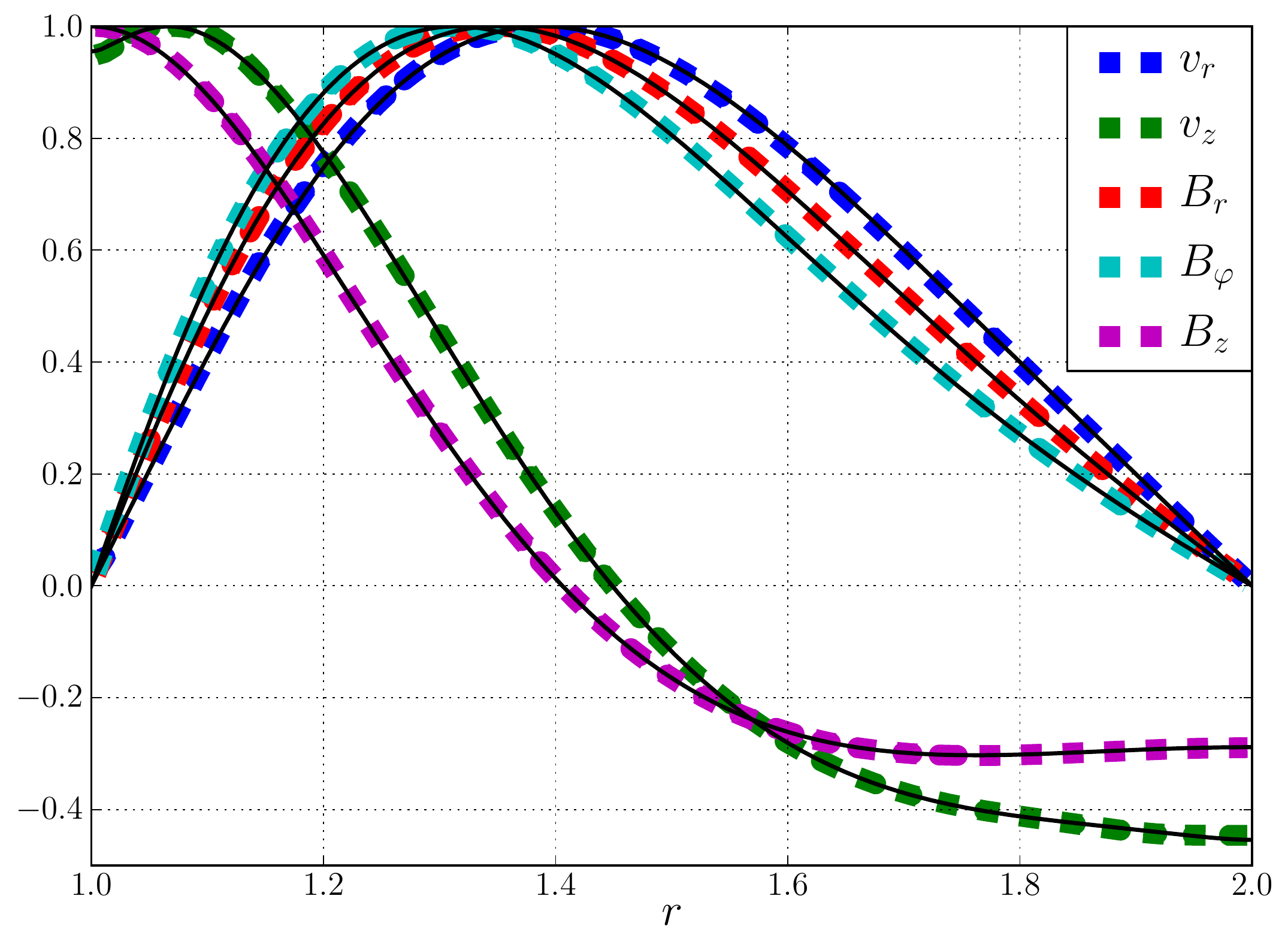}
\caption{Radial profiles of the fastest growing cylindrical MRI eigenmode for $B_0 = 2
\times 10^{-3}$, $\eta_O=3 \times 10^{-3}$ and $\lH=1.0$; comparison between
the predictions (solid lines) and DNS (squares). }
\label{fig:stabmode}
\end{figure}

We show in Fig.~\ref{fig:stabmode} a comparison of the MRI linear modes
including both ohmic diffusion and the Hall effect. This setup has $B_0 = 2
\times 10^{-3}$, $\eta_O=3 \times 10^{-3}$ and $\lH=1.0$, and would be linearly
stable in the absence of the Hall effect because of Ohmic damping with $\mathcal{S}_{\mathrm{O}} < 1$ \citep{J96}. Five of these normalized radial profiles are drawn in Fig.~\ref{fig:stabmode}. 
These profiles show that our DNS implementation accurately reproduces the spectral
predictions, with an average error of the order $10^{-3}$.

\begin{figure}[h]
\centering
\includegraphics[width=\hsize]{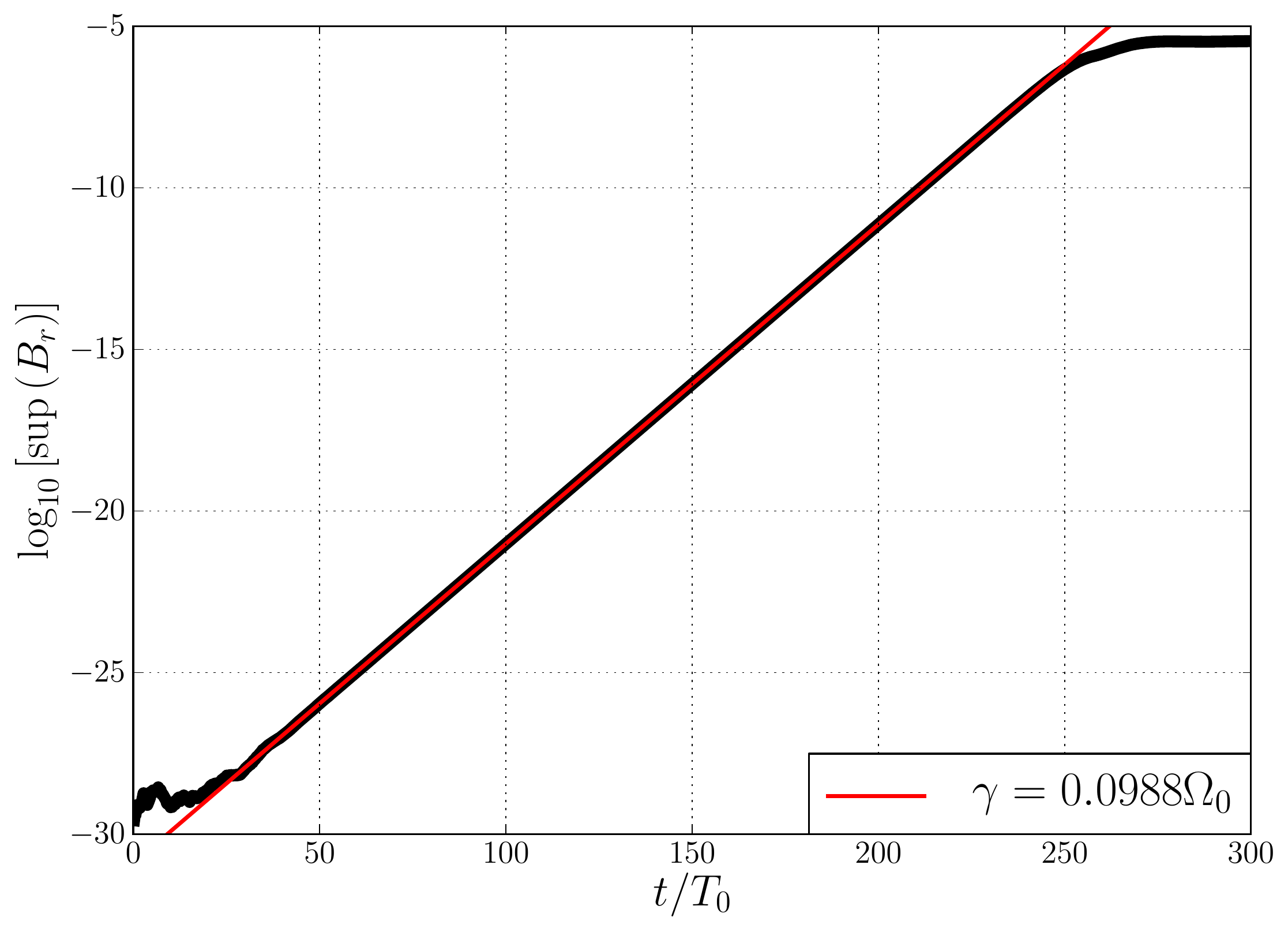}
\caption{Exponential growth of the perturbed magnetic field $B_r$. }
\label{fig:stabexpo}
\end{figure}

For this particular setup, the measured growth rate $\gamma = 0.0988$ matches
the predicted rate $\gamma_0 = 0.0989$ (see Fig.~\ref{fig:stabexpo}). In all
our tests, the predicted growth rate fit to better than $10^{-3}$ in relative
accuracy with measurements from DNS. The same linear analysis was performed in
3D cylindrical coordinates with only one computational grid cell in the
azimuthal direction, so that we preserve the axisymmetric conditions while
testing the 3D implementation of the code. Finally, we ensured that the 2D and
3D implementations gave similar results in the non-linear phase for similar
initial conditions. 

\subsection{Comparison with previous works} \label{sec:comparison}

The only global simulations including the Hall effect published to date are
those of \cite{KD14}. Their setup being similar to ours, it it is natural to
begin our report with a comparative study. We wish to reproduce their highest
resolution simulations for the two orientations of the initially axial magnetic
field with respect to the rotation axis: Up for aligned (corresponding to run
`res4-mf') and Down for oppositely directed (corresponding to their run
`480-mf-minusbz').

We first convert their units into ours and get the following values (see
appendix \ref{app:kd14}): the computational domain is $\left( r, \varphi ,z
\right) \in \left[1,5.2\right] \times \left[0,\pi/2\right] \times
\left[0,0.39\right]$, meshed with a grid $480 \times 480 \times 36$. The
isothermal sound speed is fixed to $c_s = 4.35 \times 10^{-2}$, the initial
magnetic field is $B_0 = 2.2 \times 10^{-3}$ and the Hall length is $\lH = 5.5
\times 10^{-3}$. We integrate the dynamical equations on a longer time interval
of 130 inner orbits in order to see if the total magnetic energy grows
exponentially as seen by \cite{KD14} on their figure 13. They also report the
appearance of transient waves exciting resonant modes from the radial
boundaries, motivating the inclusion of damping buffer zones. We used similar
buffers of width $0.2 r_0$ where the hydrodynamical variables $\rho$ and
$\bm{v}$ are relaxed to their initial values on a characteristic time of one
local orbital period. In addition, we chose to maintain the total vertical
magnetic flux and mass constant within the active domain, using the procedure
described in \ref{sec:gridcond} both for $B_z$ and $\rho$. This modification of
the original setup was necessary in order to keep a quasi-steady turbulent
state and measure a statistically meaningful difference between the two runs ;
otherwise, mass and flux losses where about twice larger in the first stage of run Up compared to
run Down, after which both runs had a compatible level of turbulent stress slowly decreasing with time

\begin{figure}[h]
\centering
\includegraphics[width=\hsize]{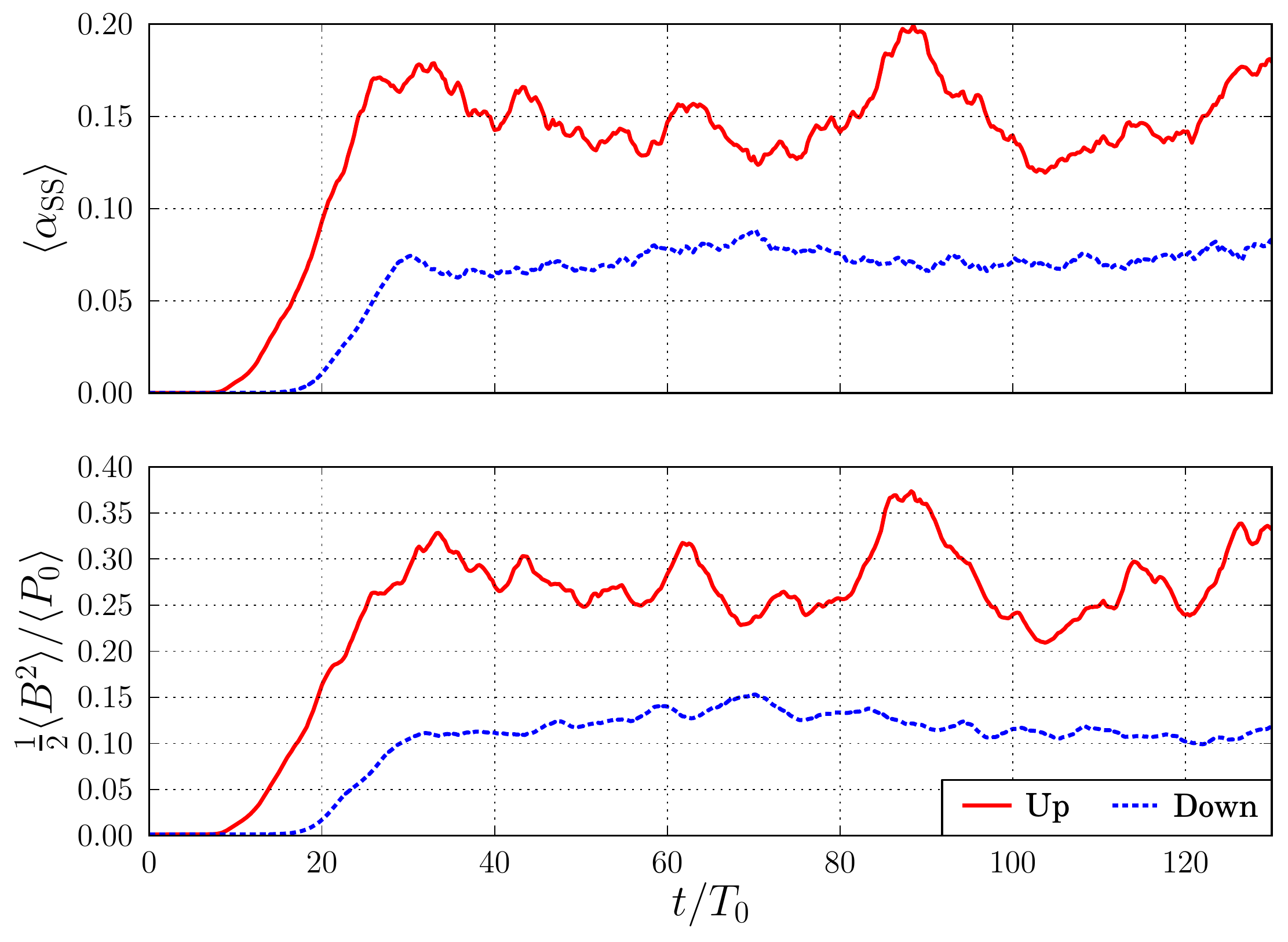}
\caption{Space-averaged $\alpha_{\mathrm{SS}}$ parameter (upper panel) and
total magnetic energy (lower panel) over time, in the active domain for the two
comparison runs Up (solid red) and Down (dashed blue). }
\label{fig:irish}
\end{figure}

We show in Fig.~\ref{fig:irish} the evolution in time of the space-averaged
$\brac{\alpha_\mathrm{SS}}$ and the total magnetic energy for the two runs,
corresponding to the figures 12 and 13 from \cite{KD14}. 

We find a slightly lower linear growth rate in run Down, in agreement with the
fact that axisymmetric configurations are stabilised by the Hall effect when
the magnetic field is anti-aligned with respect to the rotation axis \citep{BT01}. During
the steady turbulent phase from $60 T_0$ to $130 T_0$, we measure
$\overline{\alpha_{\mathrm{SS}}} = \left(1.5 \pm 0.2 \right)\times 10^{-1}$ in
run Up and $\overline{\alpha_{\mathrm{SS}}} = \left(7.4 \pm 0.5\right) \times
10^{-2}$ in run Down, larger than the results of \cite{KD14} by a factor $2$
for run Up and a factor $8$ for run Down. However, we can compare these values
with the local Hall-MHD simulations Z2L and Z4L of \cite{SS02}, where they used
similar input parameters in extended shearing-boxes: $B_0 = \pm 2.5\times
10^{-3}$, $\lH \simeq v_A/ \Omega \approx 5.5 \times 10^{-3}$ at radius
$1.8r_0$, plus an additional resistivity such that $\Lambda_{\mathrm{O}} =
100$, large enough to impact only in a minor way the saturation level of the
turbulence. These two runs yielded $\overline{\alpha_{\mathrm{SS}}} \approx
2\times10^{-1}$ in Z2L (Up case) and $\overline{\alpha_{\mathrm{SS}}} \approx 8
\times 10^{-2}$ in Z4L (Down case), similar to our results both in magnitude
and ratio. 

Concerning the total magnetic energy, we do not observe a long-term growth as
reported by \cite{KD14}. The total magnetic energy is about twice larger in run
Up compared to run Down, scaling as the ratio of stresses in accordance
with previous studies of MRI-induced turbulence in local ideal MHD simulations
\cite[see e.g.][figure 2]{MHS15}. The mean magnetic energy density is
$1.2\times 10^{-1} \rho_0 c_s^2$ in run Down, comparable again to run Z4 of
\cite{SS02} (see their figure 1). \rev{It is possible that the long-term exponential growth found by \cite{KD14} both in turbulent stress and in magnetic energy (their figures 12 to 15) reflects a state that is not converged yet, whence our larger values for $\alpha_{\mathrm{SS}}$. }

We do observe the formation of structures such as gaps close to the inner and
outer boundaries, but we attribute them to the damping buffer zones for the
density accumulates precisely at the edge between the the inner buffer and the
active domain. This is the kind of density structures we want to avoid with our
own buffers implementation as described in section \ref{sec:gridcond}. Finally,
note that the strong turbulent activity ($\alpha_\mathrm{SS}\sim 10^{-1}$) is
due to the large geometrical thickness of the disk compared to its pressure
scale height near the inner boundary ($\Omega h\gg c_s$). For this reason,
turbulence becomes transonic and strong density waves develop. This unrealistic
feature is absent from our other 3D runs for which we enforce $\Omega h\lesssim
c_s$.


\section{Results} \label{sec:3DHall}

\subsection{Numerical protocol}

All of the global simulations to date have been computed in the
weak Hall regime ($\lH/h\ll 1$) which only moderately affect the dynamics of
the system. However, disks are known to exhibit regions with much stronger Hall
effect, in particular in the so called ``dead zone'' between 1 and 10 a.u.
\citep{LKF14}. In these regions, we expect $\lH/h$ of the order of 10 in the disk
midplane. In this section we focus on this regime and on its impact on the
global disk dynamics.

The integration time varied from one run to another depending on the steadiness of the flow, with a default
lower bound of 200 inner orbits, that is about 18 orbits at the outer radius.
The boundary conditions are those described in section \ref{sec:gridcond}. We
ensured that the purely hydrodynamic case was stable, with only faint spiral
waves coming from the inner boundary ($\delta v_r / c_s \sim \delta \rho /
\rho_0 \lesssim 10^{-4}$). 

\begin{figure}[h]
\centering
\includegraphics[width=\hsize]{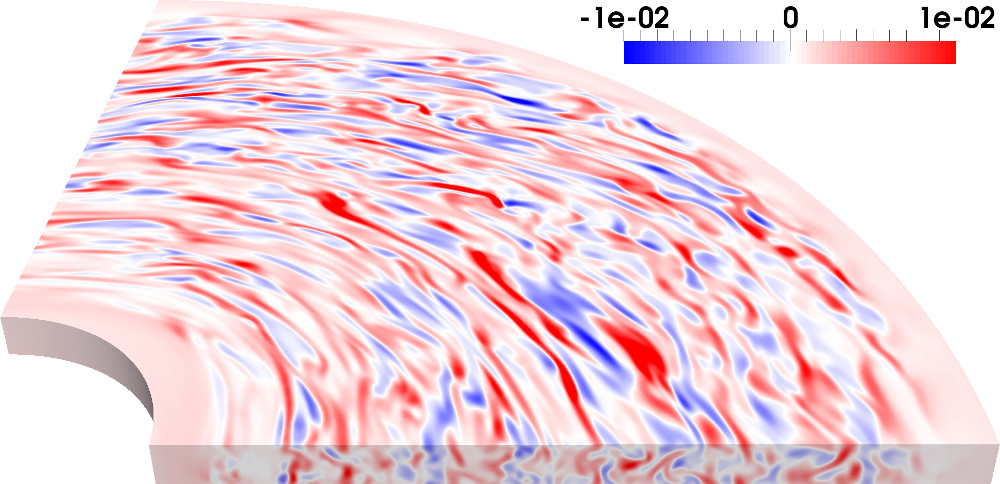}
\caption{Vertical magnetic field in the turbulent flow when switching on the
Hall effect, at $t=50 T_0$. }
\label{fig:3Dsnapturb}
\end{figure}

As we are mostly interested in the non-linear evolution of the system, we
initialise the flow in a specific way. We start from a keplerian flow in
ideal MHD, threaded by a constant vertical magnetic field $B_0 = 10^{-3}$.
Perturbations are added to the three components of the velocity field with a
large amplitude $\delta v / c_s = 10\%$, so that MRI modes rapidly develop on
the entire domain. After $t \approx 30 T_0$, the MRI saturates and we stop the
simulation at $t=50 T_0$ in a fully turbulent configuration as illustrated in
Fig.~\ref{fig:3Dsnapturb}. Only $3\%$ of the total mass has left the box at
that moment, equally from the inner and outer radial boundaries. \rev{The radial profile of $\brac{\rho}_{\varphi z}$ is decreased by about $10\%$ close to the radial boundaries, and density waves create fluctuations of about $0.1 \rho_0$ in the active domain. The turbulent state thus obtained is not subject to a radial density stratification. }

All subsequent runs are launched from this configuration, with non-ideal
effects switched on and with the net magnetic flux fixed to the desired value.
Apart from saving $50 T_0$ of computational time, this method allows us to have
the same initial conditions for all runs, and to have initial conditions that
are turbulent over the whole domain so that self-organisation processes cannot
be attributed to an excessively symmetric initial state. The parameters used in the following 3D Hall MHD runs are listed in table \ref{table:3d-hallmhd}. 

We have not systematically explored negative mean field configurations $\bm{B}_0\bm{\cdot \Omega}<0$. 
Effectively, the Hall effect is known to be sensitive to the field polarity. In particular, the vertical field 
configuration we present is known to be stable for negative weak field strengths \citep{BT01}:
\begin{align}
|v_A|<\frac{\Omega \ell_H}{2}\quad\quad\mathrm{STABILITY}.
\end{align}
This implies that sub-equipartition magnetic fields are necessarily stable in the negative polarity configuration when $\mathcal{L}=O(1)$. 
We have checked successfully that it was the case in our numerical simulations, and we won't discuss this case any further. Note however that configurations with a strong mean toroidal field can still become unstable, as demonstrated by \cite{SLK15}.

\begin{table}
Simulations with net vertical magnetic flux
\centering                          
\begin{tabular}{c c c c c c c}        
\hline\hline                 
Name & $\Delta \varphi$ & $T_s$ & $B_0$ & $\mathcal{L}$ & $\overline{\alpha}$ & state \\
\hline                        
   B3L0 & $\pi/2$ &$200$ & $10^{-3}$ & $0$ & $5.6 \times 10^{-2}$ &  turbulent \\
   B3L1 & $\pi/2$ &$200$ & $10^{-3}$ & $0.02$ & $7.8 \times 10^{-2}$ & turbulent \\	
   B3L2 & $\pi/2$ &$200$ & $10^{-3}$ & $0.04$ &$9.8 \times 10^{-2}$ & turbulent \\	
   B3L3 & $\pi/2$ &$200$ & $10^{-3}$ & $0.1$ &  $1.4 \times 10^{-1}$ & turbulent \\
   B3L4 & $\pi/2$ &$200$ & $10^{-3}$ & $0.2$ & $8.2 \times 10^{-2}$ & turbulent \\
   B3L5 & $\pi/2$ &$200$ & $10^{-3}$ & $0.4$  &  $1.6 \times 10^{-2}$ & 1 band \\	
   B3L6 & $\pi/2$ &$300$ & $10^{-3}$ & $1$ & $4.8 \times 10^{-4}$ & 4 bands \\
   2$\pi$L4 & $2\pi$ &$200$ & $10^{-3}$ & $0.2$ & $1.7 \times 10^{-1} $ & turbulence \\
   2$\pi$L5 & $2\pi$ &$400$ & $10^{-3}$ & $0.4$ & $1.8 \times 10^{-2} $ & vortex \\
   2$\pi$L6 & $2\pi$ &$200$ & $10^{-3}$ & $1$ & $1.2 \times 10^{-3} $ & 3 bands \\
   B4L0 & $\pi/2$ &$200$ & $10^{-4}$ & $0$ & $3.8 \times 10^{-3}$ & turbulent \\ 	
   B4L1 & $\pi/2$ &$300$ & $10^{-4}$ & $0.02$ & $1.7 \times 10^{-2}$ & turbulent \\ 
   B4L2 & $\pi/2$ &$200$ & $10^{-4}$ & $0.04$ &  $3.7 \times 10^{-2}$ & turbulent \\ 
   B4L3 & $\pi/2$ &$200$ & $10^{-4}$ & $0.1$ & $8.2 \times 10^{-2}$ & turbulent \\ 
   B4L4 & $\pi/2$ &$200$ & $10^{-4}$ & $0.2$ & $6.2 \times 10^{-2}$ & turbulent \\ 
   B4L5 & $\pi/2$ &$400$ & $10^{-4}$ & $0.4$ & $1.7 \times 10^{-2}$ & vortex \\ 
   B4L6 & $\pi/2$ &$300$ & $10^{-4}$ & $1$ & $3.8 \times 10^{-4}$ & 1 band \\ 
   B4L7 & $\pi/2$ &$300$ & $10^{-4}$ & $2$ & $5.1 \times 10^{-4}$ & 3 bands \\ 
   B5L1 & $\pi/2$ &$200$ & $10^{-5}$ & $0.1$ & $7.6 \times 10^{-2}$ & turbulent \\
   B5L2 & $\pi/2$ &$200$ & $10^{-5}$ & $0.2$ & $6.4 \times 10^{-2}$ & turbulent\\ 
   B5L3 & $\pi/2$ &$300$ & $10^{-5}$ & $0.4$ & $1.1 \times 10^{-2}$ & vortex \\
   B5L4 & $\pi/2$ &$300$ & $10^{-5}$ & $1$ & $4.4 \times 10^{-4}$ & 1 band \\ 
   B5L5 & $\pi/2$ &$200$ & $10^{-5}$ & $2$ & $8.5 \times 10^{-4}$ & 3 bands \\ 
   B5L6 & $\pi/2$ &$200$ & $10^{-5}$ & $4$ & $2.5 \times 10^{-4}$ & 5 bands \\
   B5L7 & $\pi/2$ &$200$ & $10^{-5}$ & $10$ & $3.7 \times 10^{-4}$ & 3 bands \\
\hline                                   
\end{tabular}
\vspace{2mm}
\caption{Parameters for the 3D Hall MHD runs: label of the run, integration
time $T_s$, mean vertical magnetic field $B_0$, Hall parameter $\mathcal{L}$,
total turbulent stress $\overline{\alpha}$, and final state of the flow. }     
\label{table:3d-hallmhd}      
\end{table}

\subsection{Turbulence}

Our study starts by sampling $\mathcal{L}$ for
different values of the global magnetic flux. We wish to find whether the
transition from high to low transport states found in KL13 still occurs
in cylindrical geometry, how it depends on the available net flux, and to
characterize these states in terms of turbulent stress and critical
$\mathcal{L}$. 

\begin{figure}[h]
\centering
\includegraphics[width=\hsize]{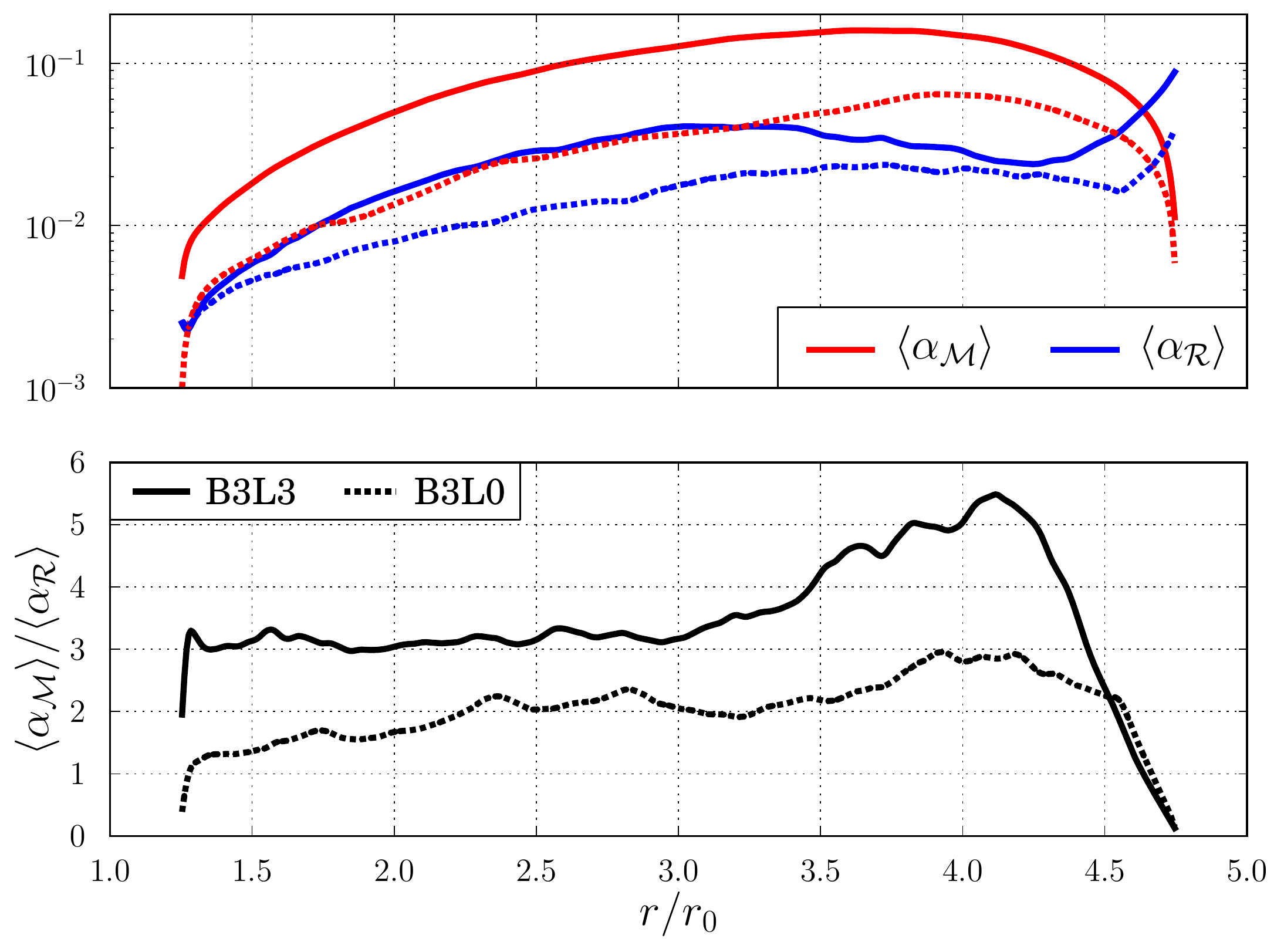}
\caption{Upper panel: radial profiles of the magnetic (red) and kinetic (blue)
contribution to $\alpha$ in runs B3L3 (solid line) and B3L0 (dashed line);
lower panel: ratio of Maxwell to Reynolds stress in the same runs. The time
averaging is performed between $100 T_0$ and $200 T_0$. }
\label{fig:stress-prof}
\end{figure}

In Fig.~\ref{fig:stress-prof} are drawn the profiles of normalized Reynolds
and Maxwell stress for the ideal-MHD run B3L0 and the Hall MHD run B3L3. Since
we kill the MRI within the buffer zones, the stress has to vanish at the
boundaries of the active domain, attaining its maximum near $3.5 r_0$. 
We note a local increase of Reynolds stress near the outer boundary, possibly
due to a characteristic damping time too long in the outer buffer. The profile of Maxwell stress is
essentially shifted of a factor two higher by the Hall effect; the Reynolds
stress does not increase as much, but we show on the lower panel that the ratio
of Maxwell to Reynolds stress profiles remains roughly constant with radius in
both runs with a value $\mathcal{M}/\mathcal{R} \approx 2$ in ideal MHD and $3$ in
Hall-MHD, comparable to previous shearing box simulations \citep[see e.g.][]{HGB95}.

\begin{figure}[h]
\centering
\includegraphics[width=\hsize]{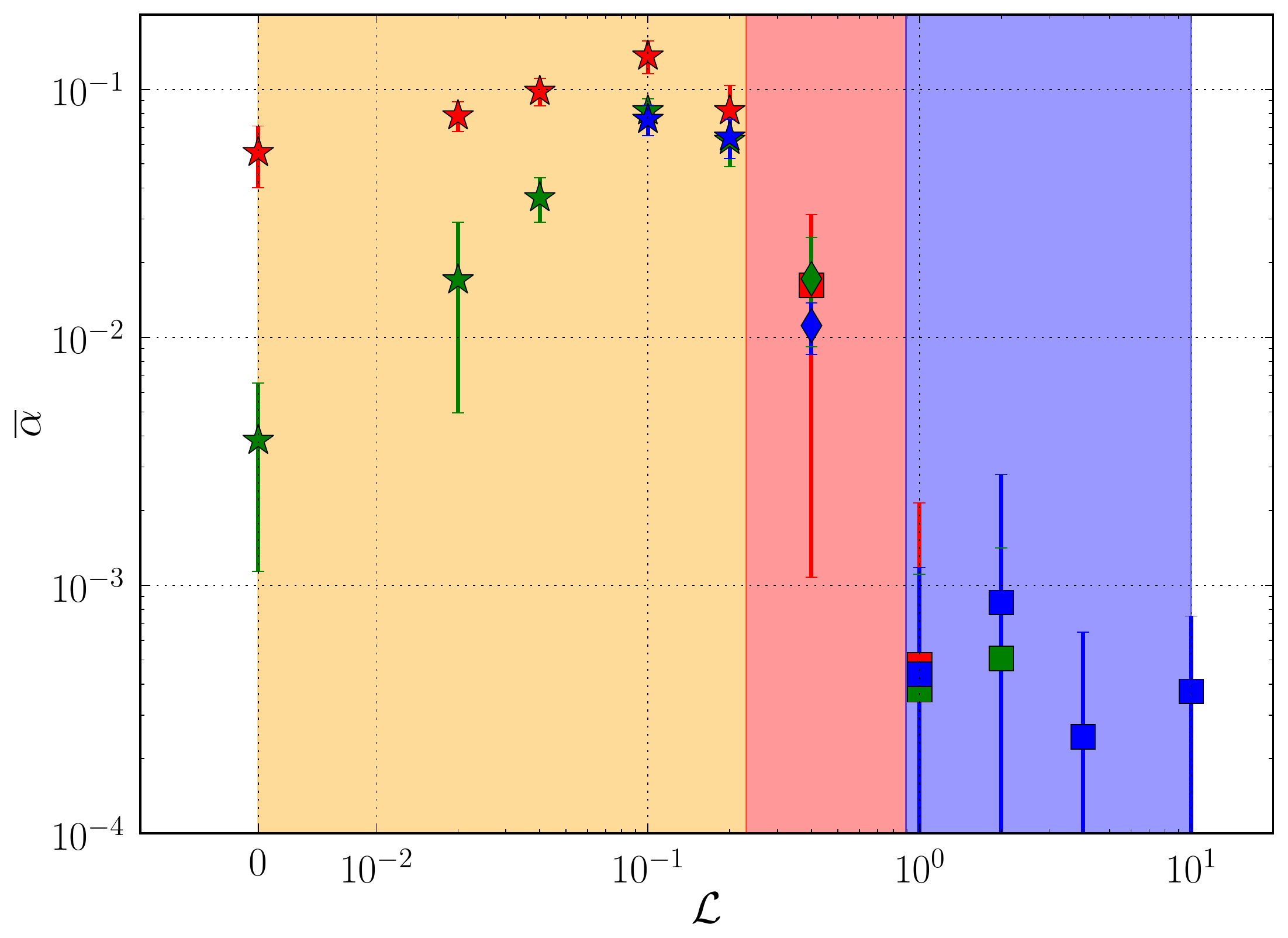}
\caption{Time averaged total turbulent stress $\overline{\alpha}$ as a function
of the Hall parameter $\mathcal{L}$ for runs B3 (red), B4 (green) and B5
(blue); the bars indicate the standard deviation over the
averaging time interval; the final state of the flow is represented by stars for turbulence (filled orange region), diamonds for vortex (filled red region) and squares for zonal flows (filled blue region). }
\label{fig:B_lh_alphat}
\end{figure}

For all the runs with $\mathcal{L} \in \left[ 0,1 \right]$, we
approximately reach a statistically steady state at the end of the simulation,
and average the global turbulent stress in time over at least $50 T_0$ ($100
T_0$ in most cases). These values of $\overline{\alpha}$ are presented in Fig.~\ref{fig:B_lh_alphat} against the corresponding value of $\mathcal{L}$. We find a good qualitative agreement with the figure 11 of KL13: the turbulent
activity is enhanced for small $\mathcal{L}$ (orange region in Fig.~\ref{fig:B_lh_alphat}), reaching its peak value for
$\mathcal{L} \approx 0.1$ at all $B_0$, beyond which the system forks to a
low-transport state with a mean stress significantly lower than in the ideal
MHD case. We also find that the stress increases with $B_0$ for
$\mathcal{L}<0.1$ in accordance with previous ideal-MHD local simulations \citep[e.g.][]{SS02}. 

A significant fraction of mass is lost in these simulations: up to $50\%$ in
run B3L3. The ratio of accreted to excreted mass is approximately $1$ in runs
B3L0 to B3L2, increases to $1.8$ in B3L3, $1.5$ in B3L4, and decreases to $0.8$
in B3L5 and only $25\%$ in B3L6. In this last run, the excreted mass is
transported by spiral density waves propagating through the entire domain, whereas
turbulent accretion is totally suppressed. The same global trend is observed
for all values of $B_0$. Despite this significant fraction of mass lost,
imposing a constant mean magnetic field made it possible to reach quasi-steady states over
several hundred inner orbits.

\subsection{Zonal flows} \label{sec:3D-zonal}

\subsubsection{Characterisation}

We move our attention to the structural properties of the system, starting with
Hall-induced zonal flows. As indicated in table \ref{table:3d-hallmhd}, we find
that the flow self-organises to long-lived axisymmetric structures whenever
$\mathcal{L} \gtrsim 1$ (blue region in Fig.~\ref{fig:B_lh_alphat}). At a given $\mathcal{L}$, the number of bands depends
on the initially available magnetic flux, more flux producing a larger number of
bands. Reciprocally, at a given $B_0$ the number of bands is found to increase
with $\mathcal{L}$, the exception being run B5L7 where the flow instantaneously
crystalized into three positive and negative magnetic field bands. 

\begin{figure}[h]
\centering
\includegraphics[width=\hsize]{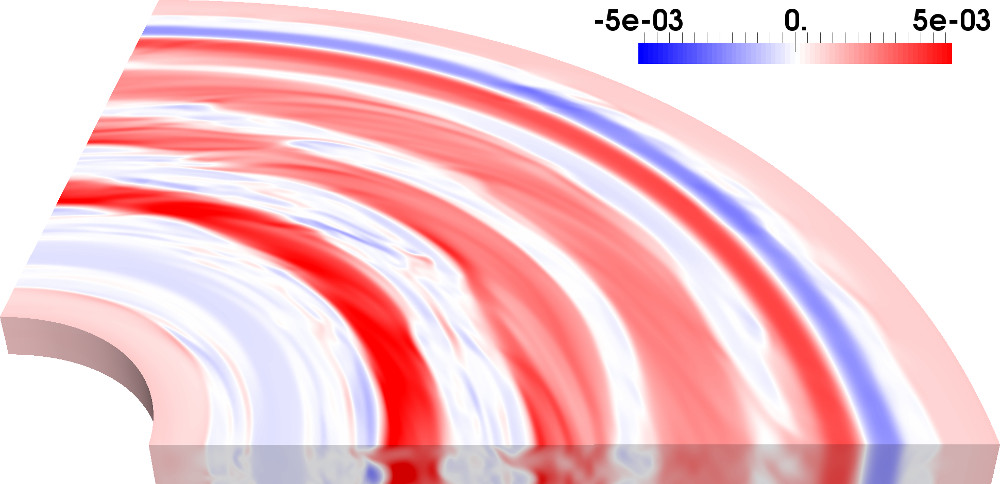}
\caption{Vertical magnetic field in the run B3L6 at $t=300 T_0$. }
\label{fig:3d-zf}
\end{figure}

We show in Fig.~\ref{fig:3d-zf} the state of run B3L6 at time $300T_0$. The
vertical magnetic field appears to be organised in four axisymmetric bands,
with little to no apparent vertical structure, and leaving the rest of the
domain almost field-free. 

\subsubsection{Physical origin} \label{sec:zforigins}

\begin{figure}[h]
\centering
\includegraphics[width=\hsize]{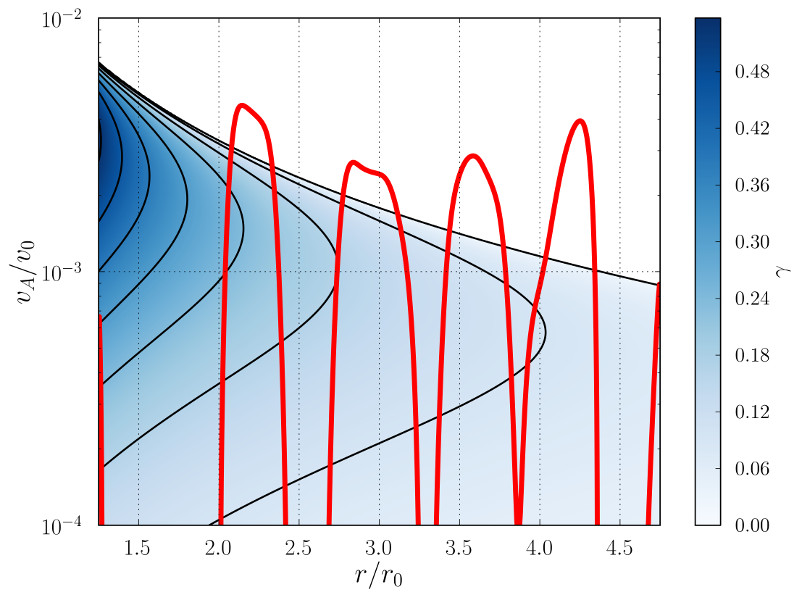}
\caption{HSI linear growth rates for the mode with vertical wave vector $k_z =
2\pi/h$ (blue color + contour lines), and local vertical Alfvén velocity (red)
in run B3L6 averaged from $200 T_0$ to $300 T_0$. }
\label{fig:3d-zf-hsi}
\end{figure}

The dynamics of these zonal fields or bands have been explored by
KL13 and we recover here a similar qualitative picture.
Let us first emphasize that the zonal field regions are dynamically stable. 
As demonstrated in Fig.~\ref{fig:3d-zf-hsi}, the vertical field strength in
each band is so strong that it quenches the linear HSI. This fact explains why 
a stronger mean initial vertical flux leads to more bands of similar width.
That being said, the bands having a limited radial extent, they should smear out radially because of
diffusion (either numerical, physical or turbulent), reducing the field
strength in the band and ultimately making the whole band HSI unstable again.
This is not observed in our simulations and we find instead a confinement
effect which calls for a proper explanation.

\begin{figure}[h]
\centering
\includegraphics[width=\hsize]{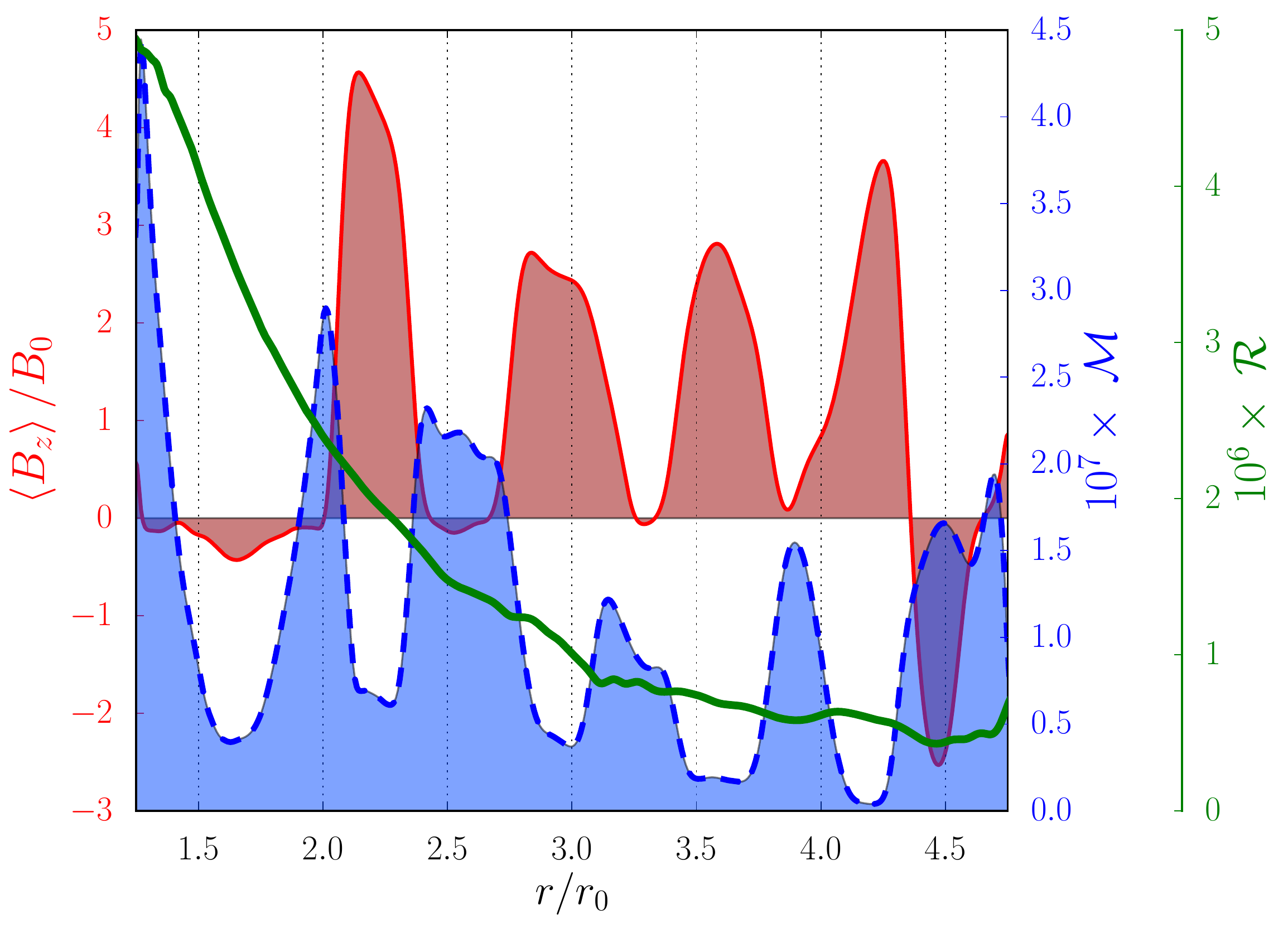}
\caption{Radial profiles of $B_z$ (filled red), $10^7 \times
\mathcal{M}_{r\varphi}$ (dashed, filled blue) and $10^6 \times
\mathcal{R}_{r\varphi}$ (solid green) in run B3L6, averaged in time between
$160T_0$ and $230T_0$. }
\label{fig:3d8bands}
\end{figure}

\begin{figure}[h]
\centering
\includegraphics[width=\hsize]{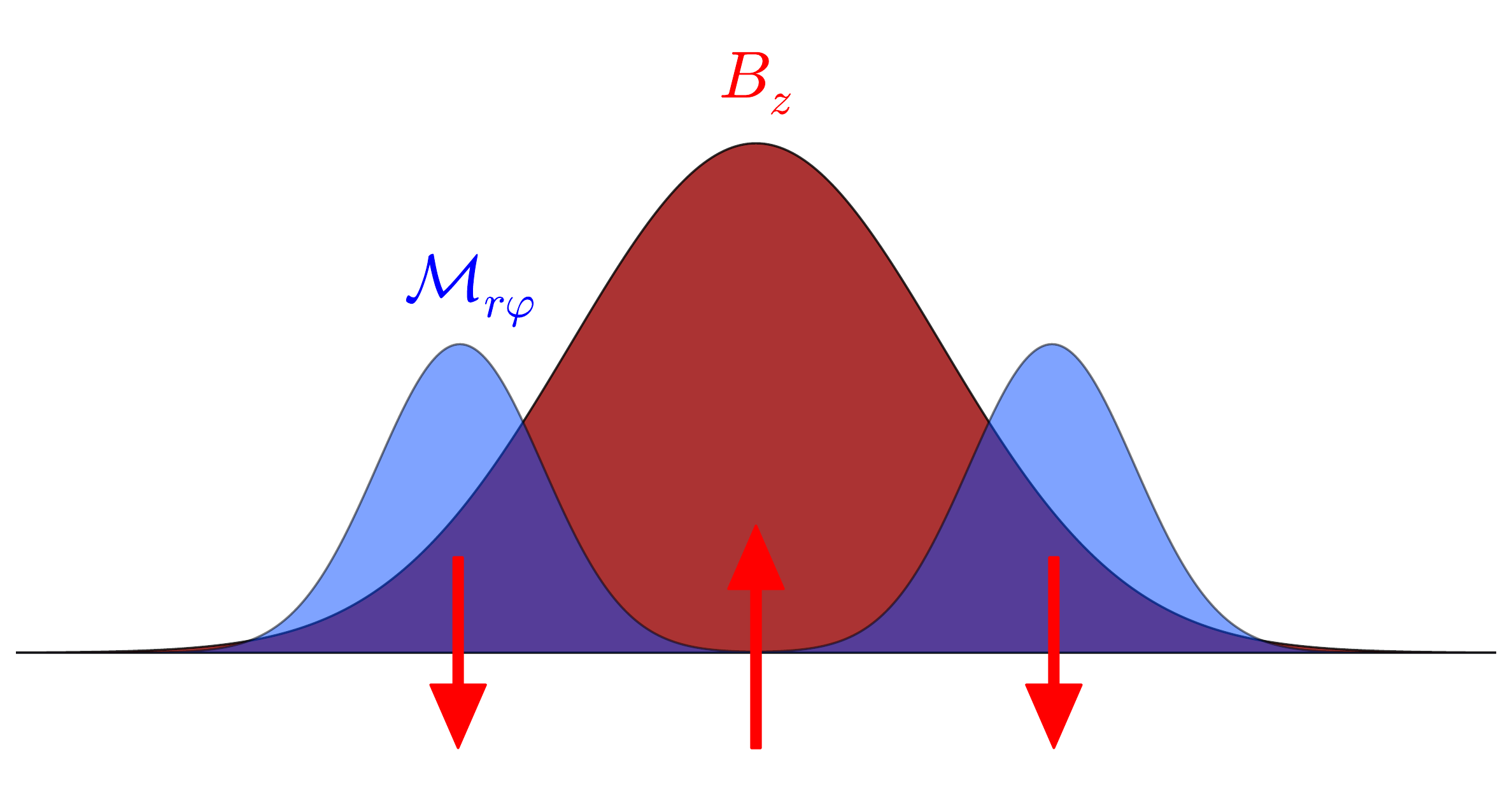}
\caption{$B_z$ confinement mechanism by belts of Maxwell stress $\mathcal{M}_{r
\varphi}$: the magnetic field decreases in regions of negative second
derivative of $\mathcal{M}$, and increases where the curvature of $\mathcal{M}$
is positive. }
\label{fig:bzmaconf}
\end{figure}

The zonal field confinement takes its roots in the HSI at the
boundary of each bands: the vertical flux is weaker but non-zero, allowing
localised HSI modes to develop. This leads to the generation of a localised
Maxwell stress at the border of each band, as illustrated in Fig.~\ref{fig:3d8bands}. However,
in Hall-MHD, the Maxwell stress appears explicitly in the induction equation \eqref{eq:inucinde}
through the Hall term. This equation predicts that a local maximum of the Maxwell stress pushes away positive
vertical flux tubes. As a result, the growth of HSI modes at the border of each band pushes the vertical magnetic flux back in the
band, thereby confining the flux in the HSI-stable region. This confinement mechanism by the Maxwell stress is sketched in Fig.~\ref{fig:bzmaconf} and explains the global organisation observed in these simulations.

The bifurcation to such a self-organised state from a fully turbulent state occurs when confinement overcomes diffusion. KL13 show from equation \eqref{eq:inucinde} that the transition happens near a critical $\mathcal{L}_0$ which depends on the stabilising Alfv\'en velocity \eqref{eq:HSIeasy} and on the turbulent magnetic Prandtl number $P_m \equiv \eta_t / \overline{\alpha} \Omega h^2$ via
\begin{equation} \label{eqn:kl13crit}
\mathcal{L}_0 \simeq \frac{1}{P_m}\frac{v_{A,\mathrm{crit}}}{\Omega h}. 
\end{equation}
Using $P_m \approx 2$ as measured in local simulations \citep{LL09}, they deduce $\mathcal{L}_0 \approx 0.2$. The same argument holds in our case, and the fact that the transition still happens near this critical value translates into a turbulent magnetic Prandtl number of order unity again in a global, non-stratified setup. 

Looking at the Reynolds component of the stress, we show in Fig.~\ref{fig:3d8bands} that
 it is the dominant component of the total stress in this organised configuration, by a factor ten above the Maxwell
stress. The physical origin of this stress resides in large-scale spiral waves, unaffected by the presence of zonal fields. We found a plateau of total stress when increasing $\mathcal{L}$ beyond unity although the Maxwell stress kept decreasing below $\overline{\alpha_{\mathcal{M}}} < 10^{-4}$ near the end of the simulation. After ruling out a possible defect of our inner damping regions, we think these waves may be excited by residual turbulence and local minima of potential vorticity near the zonal flows.

\subsection{Vortices}

\subsubsection{Characterisation}

\begin{figure}[h]
\centering
\includegraphics[width=\hsize]{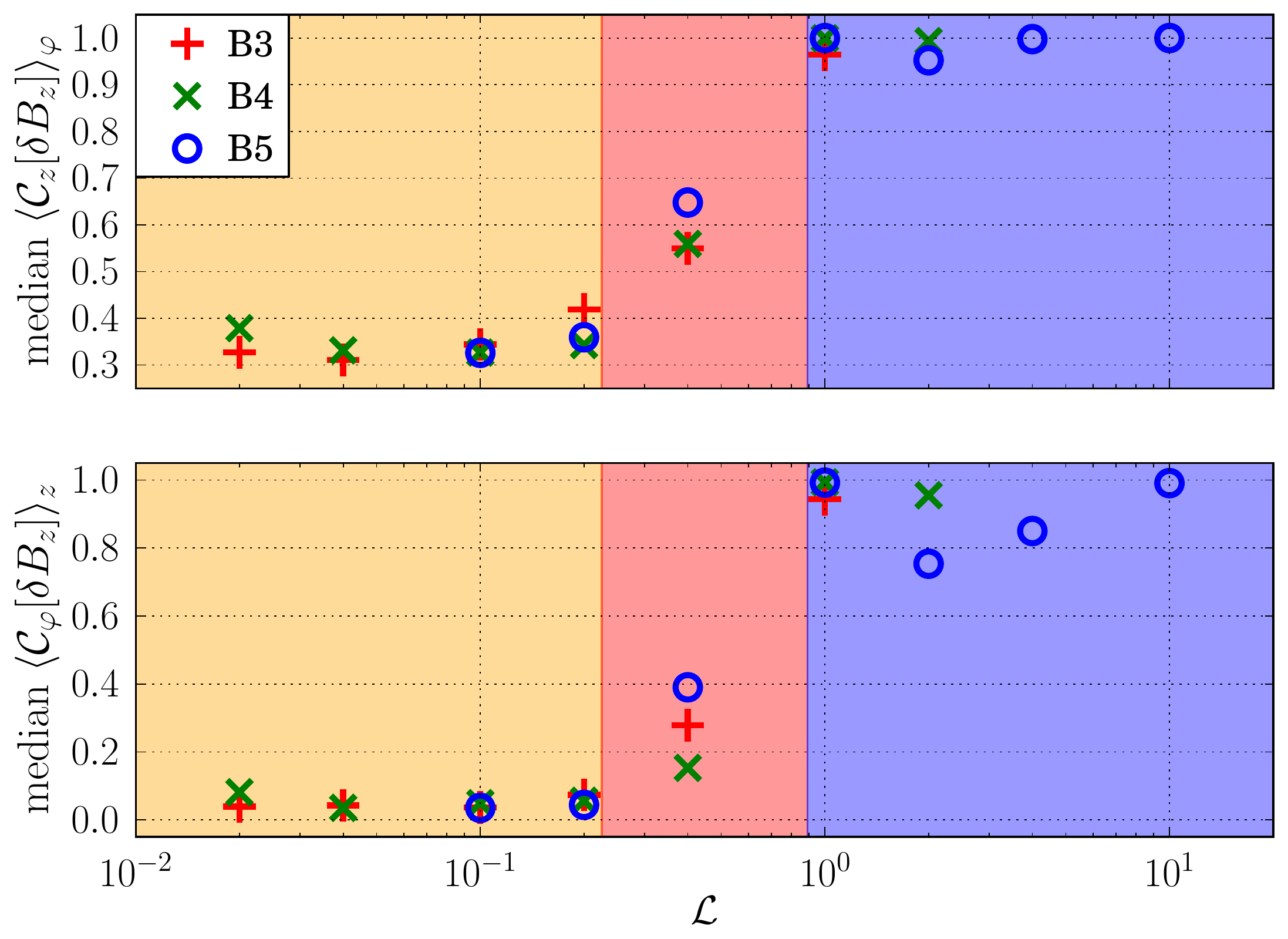}
\caption{Median value of the vertical (upper panel) and azimuthal (lower panel)
auto-correlation profiles of the vertical magnetic field $\delta B_z$, measured
at the end of the quarter-disk runs B3 (red plus), B4 (green cross) and B5 (blue circles); the orange, red and blue regions correspond respectively to a turbulent, vortex and zonal flow final state. }
\label{fig:corryz}
\end{figure}

For intermediate Hall strengths, the level of turbulent stress
remains quite high but the aspect of the turbulent structures gets more clumpy
as $\mathcal{L}$ is increased (red region in Fig.~\ref{fig:B_lh_alphat}). We quantify this structural transition in
Fig.~\ref{fig:corryz}, with the median value (over the radial extent of the
disk) of the normalized correlation lengths of $\delta B_z \equiv B_z - B_0$ in the
vertical and azimuthal directions. The change from small to large-scale
fluctuations occurs rapidly for $\mathcal{L} \gtrsim 0.2$, and when $\mathcal{L}
\geq1$ the correlation factor is stalled near unity. 

\begin{figure}[h]
\centering
\includegraphics[width=\hsize]{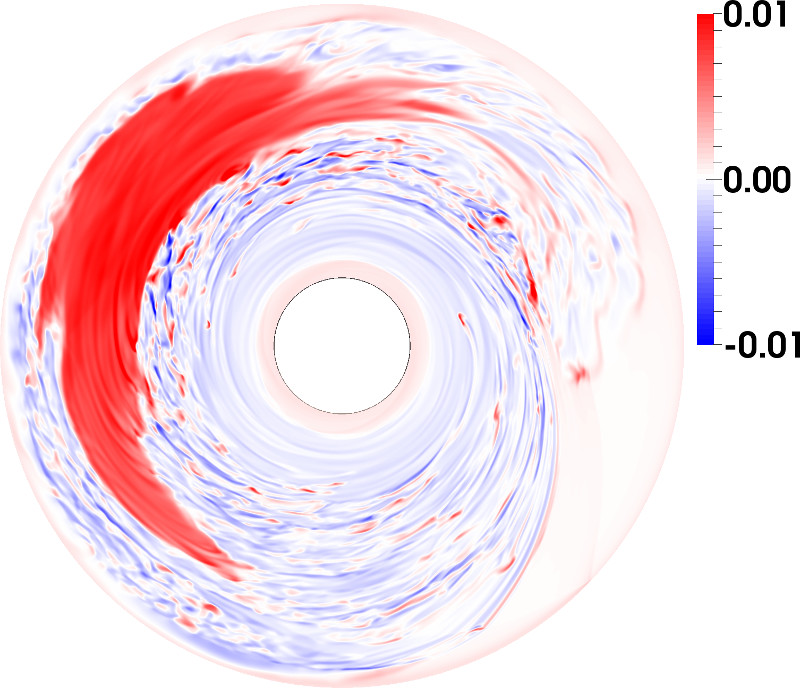}
\caption{Vertical magnetic field in run 2$\pi$L5 at time $300 T_0$. }
\label{fig:snapvbz}
\end{figure}

During the transition $\mathcal{L} \approx 0.4$, the turbulent fluctuations
merge to form a sustained patch of magnetic field in runs B4L5 and B5L5. One band was formed in run B3L5 with the same $\mathcal{L}$; Fig.~\ref{fig:B_lh_alphat} shows that it is affected by large fluctuations in stress over time\rev{, indicating that the structure is not steady}. In order to test its stability, we ran the full-disk simulation 2$\pi$L5 and found that the band would actually break and rather form a large patch as illustrated in Fig.~\ref{fig:snapvbz}. In this run the average vertical field is $B_z \approx 8 \times 10^{-3}$ inside the patch. 
Similarly to the zonal flows of section \ref{sec:3D-zonal}, this region displays no structure in the vertical direction.  
We verified that these magnetic islands were not a mere product of the global flux adjustment procedure
(see \ref{sec:gridcond}). These large scale patches have not been observed in KL13 local simulations probably because their horizontal extension covers several geometrical scale height $h$ in radius and azimuth. \rev{For larger $\mathcal{L} \gtrsim 1$, the bands are stable in the full-disk configuration as confirmed in run 2$\pi$L6. The formation of a band in run B3L5 is facilitated by the smaller angular extent of the domain, and therefore accidental. We also confirm with run 2$\pi$L4 that lower values for $\mathcal{L} \lesssim 0.2$ keep the flow turbulent in $2\pi$ disk simulations. } 

\begin{figure}[h]
\centering
\includegraphics[width=\hsize]{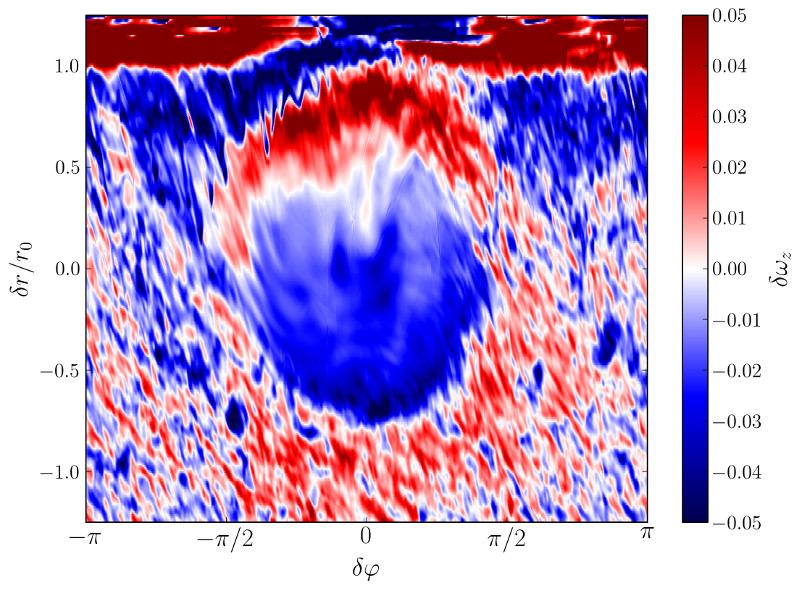}
\caption{Vertical vorticity fluctuation $\delta \omega_z$ in the $\left( \varphi , r \right)$ plane of run 2$\pi$L5, centered on the vortex, averaged in the vertical direction and in time with five snapshots between $250T_0$ and $290T_0$; the vertical axis $\delta r / r_0$ is the radial distance to the measured center of the vortex. }
\label{fig:V3vortz}
\end{figure}

In the limit of incompressible Hall-MHD, the canonical vorticity 
\begin{equation} \label{eqn:vortcan}
\bm{\varpi} \equiv \nabla \times \bm{v} + \bm{B} / \rho \lH
\end{equation}
is a conserved quantity (see equation 7 of KL13); an increase in magnetic flux therefore comes with a decrease in vorticity flux. We show in Fig.~\ref{fig:V3vortz} the vertical component of the vorticity fluctuation from the initial keplerian flow: $\delta \bm{\omega} \equiv \nabla \times (\bm{v}-r^{-1/2} \bm{e_{\varphi}})$, averaged in time over $40$ inner orbits for better discernibility. We observe that the accumulation of magnetic flux is indeed balanced by a decrease in vorticity flux: $\delta \omega_z \simeq -\delta B_z /\rho \lH \approx 3 \times 10^{-2}$, making this patch a true vortex and attesting the role of the Hall effect in its formation. Its center is localised near $3.8r_0$, and the ratio of its major to minor axis is $\chi \approx 5.2$; the corresponding proper rotation period is
\begin{equation}
\delta t \simeq \frac{2\pi}{\delta \omega_z} \frac{1+\chi^2}{\chi} \approx 180 T_0,
\end{equation}
or approximately $25$ local orbits. This turn-over time is large compared to the local dynamical time-scale, which is why our time averages suffer from random fluctuations.

\subsubsection{Confinement mechanism}

The radial confinement of the vortex obeys the same mechanism as for the zonal flows; we can address it with a one-dimensional approach similar to the axisymmetric model of section \ref{sec:simo}. In the case of zonal flows, a band of magnetic field is pushed from both its inner and outer sides by the $(r,\varphi)$ component of the Maxwell stress $\mathcal{M}_{r\varphi}$, which is the product of the magnetic field component along the streamlines $B_{\varphi}$ with the component perpendicular to the streamlines $-B_r$. A vortex is defined by closed streamlines, dragging and wrapping the horizontal magnetic field around. In a frame following the vortex at its local keplerian velocity, one can use the magnetic field components along ($B_{\parallel}$) and perpendicular ($-B_{\perp}$) to the velocity streamlines. The product of these components defines a projected Maxwell stress $\mathcal{M}_{\perp \parallel}$; this stress appears at the boundary of the magnetic patch and pushes magnetic flux in the perpendicular direction $-\bm{e_{\perp}}$, thus confining the vortex from all directions. 

\begin{figure}[h]
\centering
\includegraphics[width=\hsize]{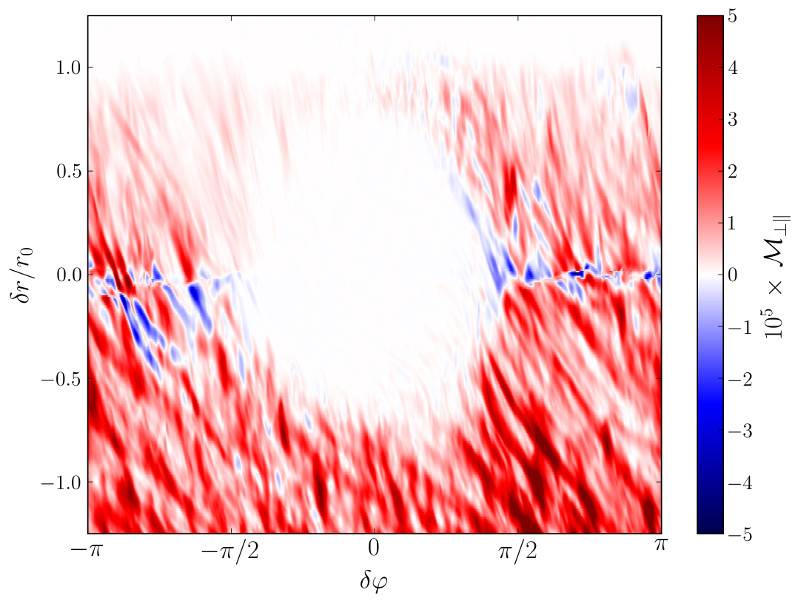}
\caption{Projected stress in the $\left( \varphi , r \right)$ plane of run 2$\pi$L5, centered on the vortex, averaged in the vertical direction and in time with five snapshots between $250T_0$ and $290T_0$; the vertical axis $\delta r / r_0$ is the radial distance to the measured center of the vortex. }
\label{fig:stackedmab}
\end{figure}

We confirm this picture by showing the averaged map of projected Maxwell stress $\mathcal{M}_{\perp \parallel}$ of run 2$\pi$L5 in Fig.~\ref{fig:stackedmab}. We find that it is negligibly small inside the vortex, attesting of the MHD stability of this region, while the overall positive stress around the vortex provides the required confinement.


\section{\label{sec:dust}Dust trapping}

An important issue is the ability of the previous structures to accumulate dust
particles in the process of planetary formation. It is known that, due to aerodynamic drag,
dust grains migrate outwards in super-keplerian regions and migrate inwards in sub-keplerian regions
\citep{W77}. A disc region trapped between a super-keplerian part at its inner side and a sub-keplerian
part at its outer side therefore constitute a dust trap\footnote{These regions are sometimes refered to as pressure bumps. As a result of the radial geostrophic equilibrium, the trap we describe necessarily correspond a to pressure bump, but we prefer avoiding this terminology since dust grains are only sensitive to the gas velocity, and not to the gas pressure. \label{fnt:dust}} where dust grains accumulate. 

In the incompressible Hall-MHD limit, an increase in magnetic flux should come with a decrease in vorticity flux, so that the flux of canonical vorticity defined by equation \eqref{eqn:vortcan} is conserved. This is equivalent to a velocity profile steepened in regions of accumulated magnetic field and flattened outside, making both zonal flows and vortices potential dust traps.

\subsection{Capture by zonal flows}

\begin{figure}[h]
\centering
\includegraphics[width=\hsize]{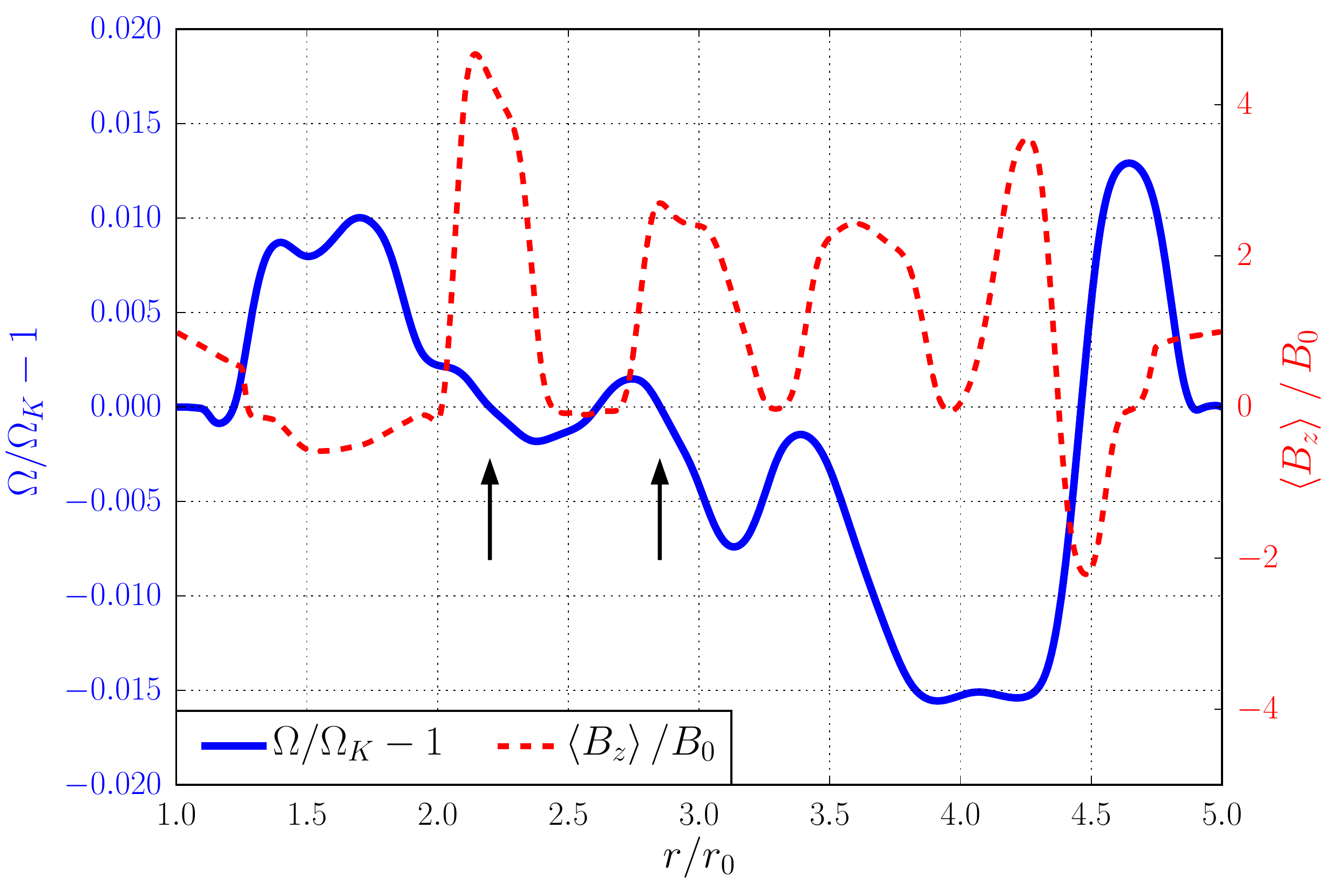}
\caption{Relative fluctuations of angular velocity (solid blue) and vertical
magnetic field (dashed red) in run B3L6, averaged in time between $260T_0$ and
$300T_0$. The arrows indicate regions of favored dust accumulation. }
\label{fig:zf-om}
\end{figure}

To see how the conservation of the canonical vorticity flux is altered in our compressible case, we draw in
Fig.~\ref{fig:zf-om} the deviation from a keplerian rotation profile in run
B3L6. We observe a transition from super to sub-keplerian rotation speed only in the
first two bands. In the outer half, the initially turbulent state plus the steady
mass excretion by spiral waves have reduced the average density to about $10\%$
of its initial value and slightly increased it in the center of the radial
domain; this global density (and therefore pressure) gradient is sufficient to
maintain a sub-keplerian flow in the outer part of the disk, preventing dust
trapping in the last two bands. 

In general, protoplanetary disks are expected to be globally sub-keplerian
thanks to a mean negative pressure gradient\footnote{This aspect has not been
included in our non-stratified model.}. This deviation from keplerian rotation can be computed
by considering the radial equilibrium
\begin{align}
\frac{\Omega}{\Omega_K}=\Bigg(1+\frac{\partial_r P}{\rho R
\Omega_K^2}\Bigg)^{1/2}.
\end{align}
The average pressure gradient entering this equation can be estimated from the
aspect ratio of the disk $\varepsilon=c_s/R\Omega_K$ so that one expects 
\begin{align}
\frac{\Omega}{\Omega_K}-1\simeq -\frac{1}{2}\varepsilon^2.	
\end{align}
Assuming a typical value of $\varepsilon=0.1$, we find that the deviations to
Keplerian rotation driven by our zonal flows in Fig.~\ref{fig:zf-om} are
marginally sufficient to create super-keplerian regions (and thus dust
traps) when this mean pressure gradient is included. For this reason, we
expect that only a few of these zonal flows (the strongest ones) can trap dust 
in models including a realistic mean pressure gradient. 

\subsection{Capture by vortices}

The possible role of hydrodynamic vortices as dust traps in protoplanetary disks was highlighted by \cite{BS95} and \cite{TBDP96}, and has now received both analytical and numerical investigations \citep[e.g.][]{JAB04,C00} . We focus on the necessary condition for a vortex to be a dust trap: that it is over-pressured with respect to the surrounding flow (see footnote \ref{fnt:dust}). In our globally isothermal simulations, this translates into an over-density in the vortex. 

\begin{figure}[h]
\centering
\includegraphics[width=\hsize]{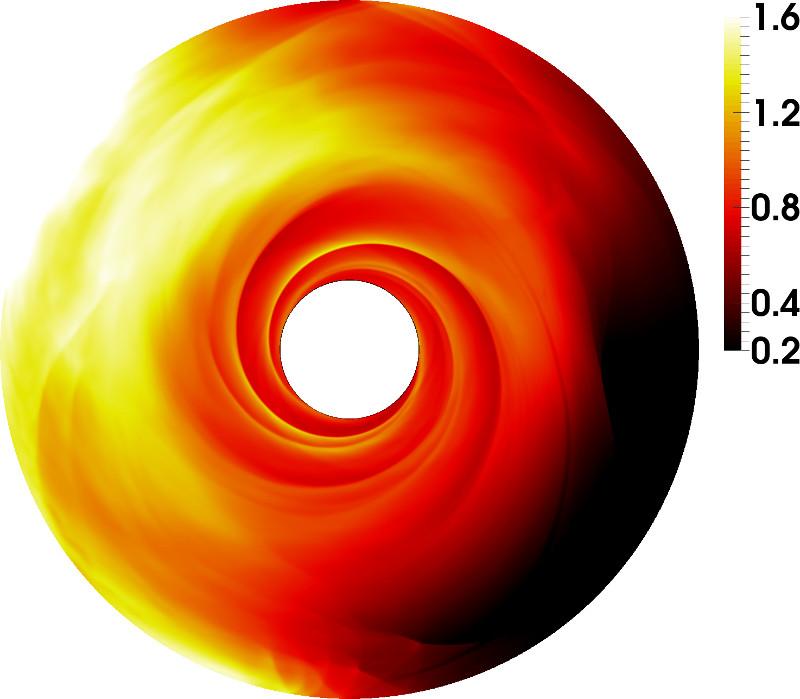}
\caption{Density distribution in run 2$\pi$L5 at time $300 T_0$. }
\label{fig:2pirho}
\end{figure}

We show in Fig.~\ref{fig:2pirho} a snapshot of the density distribution in run 2$\pi$L5 at time $300T_0$. Comparing with  Fig.~\ref{fig:snapvbz}, there is clearly an over-density at the location of the vortex, with mean value $1.5\rho_0$ inside at $300T_0$, so this vortex would be able to trap dust particles.


\section{\label{sec:threats}Threats to self-organisation}

The simulations presented so far were done in the Hall MHD regime with an
imposed mean vertical magnetic field. However, stratified shearing box models
\citep[e.g.][]{LKF14} also indicate that the magnetic field could be
essentially toroidal in the midplane of protoplanetary disks, which could in
turn affect the formation and stability of zonal flows. Besides, ionisation
models \citep[e.g.][]{SLK15} suggest that the Hall effect is dominant only in
intermediate regions ($1\,\mathrm{a.u.}\lesssim r\lesssim 30\,\mathrm{a.u.}$)
of protoplanetary disks, the inner and outer region being respectively
dominated by Ohmic and ambipolar diffusion \citep{TFG14}. These diffusive
processes could prevent the formation of sharp magnetic accumulations. 
We investigate the robustness of our previous findings against these effects in
the three next sections.

\subsection{Toroidal field}

\subsubsection{Method}

We ran eight simulations where both the vertical and azimuthal magnetic fluxes
are imposed and adjusted at every timestep. As before, we let a disk evolve
over $50 T_0$ from a keplerian flow with initial magnetic field $\bm{B} = B_0
\bm{e_z} + B_{\varphi} \bm{e_{\varphi}}$ to a fully turbulent flow, where
$B_0=B_{\varphi}=10^{-3}$ initially constant over the whole domain. This choice
of initial conditions does not correspond to an MHD equilibrium, but the intensity of the
magnetic field is weak enough to reach a steady turbulence in an overall
keplerian flow. We activated the Hall effect from this starting point in all
eight runs, and lowered the average vertical magnetic field to $B_0=10^{-4}$
for half of them, keeping $B_{\varphi}=10^{-3}$. The stopping time is set to
$300 T_0$ for runs T3L3 and T4L2, and to $200 T_0$ for the others. 
The parameters of these runs are given in Table \ref{table:3d_tor}. 

\subsubsection{Results}

\begin{table}      
Simulations with net vertical and toroidal magnetic flux
\centering                          
\begin{tabular}{c c c c c}        
\hline\hline                 
Name & $B_0$ & $\mathcal{L}$ & $\overline{\alpha}$ & state \\    
\hline                        
   T3L0 & $10^{-3}$ & $0$ & $6.2 \times 10^{-2}$ &
turbulent \\
   T3L1 & $10^{-3}$ & $0.1$ & $1.4 \times 10^{-1}$ &
turbulent \\	
   T3L2 & $10^{-3}$ & $0.4$ & $1.0 \times 10^{-2}$ &
vortex \\	
   T3L3 & $10^{-3}$ & $1$ & $1.1 \times 10^{-3}$ & 3 bands \\
   T4L0 & $10^{-4}$ & $0$ & $4.9 \times 10^{-3}$ &
turbulent \\
   T4L1 & $10^{-4}$ & $0.1$ & $7.9 \times 10^{-2}$ &
turbulent \\	
   T4L2 & $10^{-4}$ & $0.4$ & $7.3 \times 10^{-3}$ &
turbulent \\
   T4L3 & $10^{-4}$ & $1$ & $3.3 \times 10^{-3}$ & 1 band \\ 	
\hline                                   
\end{tabular}
\caption{Parameters for the 3D Hall MHD runs with toroidal field: label of the
run, mean vertical magnetic field $B_0$, Hall parameter $\mathcal{L}$,
total averaged turbulent stress $\overline{\alpha}$ and final state. } 
\label{table:3d_tor}      
\end{table}

From the last column of Table \ref{table:3d_tor}, we notice minor differences
in the final state of these simulation compared to the case of a purely
vertical mean field: run T3L3 produced 3 bands instead of 4, and run T4L2 did
not produce a distinct vortex but only short-lived patchy magnetic islands.

\begin{figure}[h]
\centering
\includegraphics[width=\hsize]{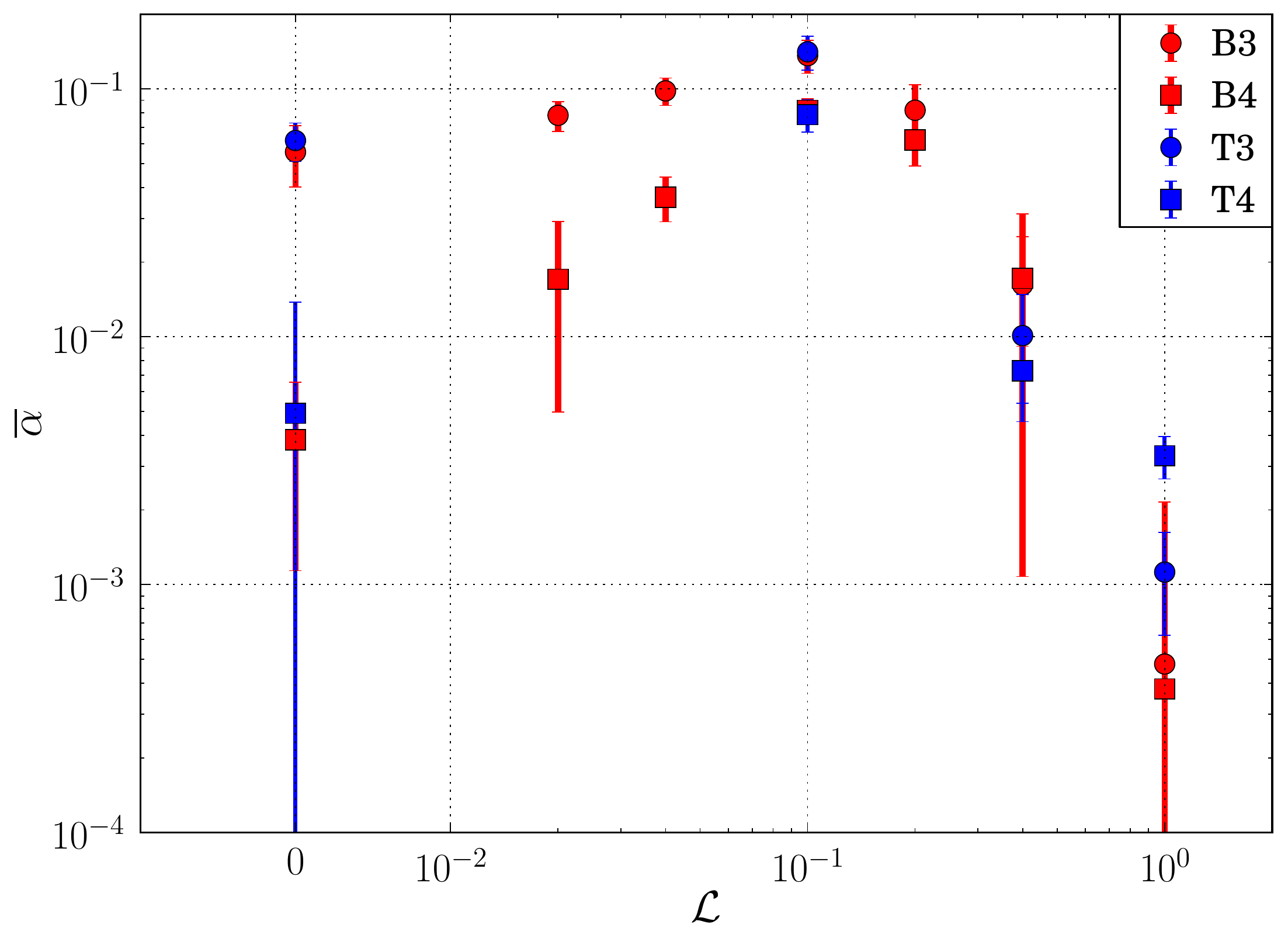}
\caption{Time-averaged stress $\overline{\alpha}$ as a
function of the Hall parameter $\mathcal{L}$, with and without a
toroidal field (blue and red respectively), with mean vertical field $B_0 =
10^{-3}$ (circles) and $B_0 = 10^{-4}$ (squares); the bars represent standard deviations of $\brac{\alpha}$ over time. }
\label{fig:BTalpha}
\end{figure}

We then compare the average stress in these toroidal runs with the previous vertical ones in Fig.~\ref{fig:BTalpha}. 
The addition of a toroidal magnetic field is seen to have no significant impact on the stress. 
In particular, the transition from high to low transport states near $\mathcal{L}\approx 0.1$ is preserved.
Only the run T4L3 shows a turbulent stress significantly higher than the analogue run with zero net toroidal flux. 
This stress comes essentially from the Reynolds component: in all the runs with
$\mathcal{L}=1$ the Maxwell stress kept decreasing below $\overline{\alpha_{\mathcal{M}}}<10^{-4}$ at the end of the
simulation, but since the Reynolds stress was statistically steady at a much higher values we did not prolong these runs further. 
 
These differences attest the presence of a slightly stronger turbulent activity, able to weaken
magnetic structuring but not drastically changing the picture. 

\subsection{Ohmic diffusion}

\subsubsection{Linear stability}

The presence of ohmic resistivity narrows the range of accessible HSI modes. In the strong Hall limit, the marginal stability of axisymmetric Hall-shear modes with wave vector $k$ in a background vertical field requires \citep{BT01} :
\begin{equation} \label{eqn:dispohm}
\lH v_A \left( \lH v_A k^2 - q \Omega \right) + \eta_{\mathrm{O}}^2 k^2 = 0. 
\end{equation}
The addition of resistivity affects the two bounds between which instability occurs :
\begin{equation} \label{eqn:ohmstab}
\frac{2\lH v_A k^2}{q\Omega} = 1 \pm \sqrt{1 - \left( \frac{2 \eta_{\mathrm{O}} k^2}{q\Omega}\right)^2}. 
\end{equation}
In order to study self-organisation, the system must initially be between these two bounds. A minimum magnetic field intensity is necessary for linear instability and the possibility of sustaining turbulence. In the other limit, the field intensity required to damp the largest scales $k_0=2\pi/h$ is lowered by diffusivity, which may help the formation of Hall-shear stable regions and favor self-organisation. We tune this range of unstable magnetic fields with the magnetic Reynolds number $\mathcal{R}_{\mathrm{O}} \equiv q \Omega / 2 \eta_{\mathrm{O}} k_0^2$.

\subsubsection{Method}

The numerical setup used in this section is the same as in section
\ref{sec:3DHall}, but with a constant resistivity $\eta_{\mathrm{O}}$ in the
active domain (and always the resistivity $\eta_b$ in the damping buffer zones). 
We sampled two values of magnetic Lundquist numbers $\mathcal{S}_{\mathrm{O}}=\{2.5,25\}$, and the magnetic Reynolds number $\mathcal{R}_{\mathrm{O}}$ is appreciably close to unity only in the small $\mathcal{S}_{\mathrm{O}}=2.5$ case. These values should be sufficiently low to produce observable effects in our runs, but not prevent self-organisation by immediately damping any perturbation. All four runs are integrated in time from $50T_0$ to $200 T_0$. 

\subsubsection{Results}

\begin{table}
\centering  
Simulations with ohmic diffusion\\
\vspace{0.2cm}
\begin{tabular}{c c c c c c}        
\hline\hline                 
Name & $B_0$ & $\mathcal{S}_{\mathrm{O}}  \:(\mathcal{R}_{\mathrm{O}})$ & $\mathcal{L}$ &
$\overline{\alpha}$ & state \\    
\hline                        
   O3N0 & $10^{-3}$ & $25 \:(119\Omega)$ & $0.4$ & $1.2 \times 10^{-2}$ &  vortex \\
   O3N1 & $10^{-3}$ & $25 \:(119\Omega)$ & $1$ & $5.2 \times 10^{-4}$ & 2 bands \\	
   O3N2 & $10^{-3}$ & $2.5 \:(12\Omega)$ & $1$ &$1.2 \times 10^{-3}$ & 1 band \\	
   O4N3 &$10^{-4}$ & $2.5 \:(119\Omega)$ & $1$ & $9.6 \times 10^{-4}$ & 1 band \\
\hline                                   
\end{tabular}
\vspace{2mm}
\caption{Parameters for the 3D Hall+Ohm MHD runs with toroidal field: label of
the run, mean vertical magnetic field $B_0$, magnetic Lundquist numbers
$\mathcal{S}_{\mathrm{O}}$ and corresponding Reynolds number $\mathcal{R}_{\mathrm{O}}$, Hall parameter $\mathcal{L}$, turbulent stress $\overline{\alpha}$ and final state of the flow. } 
\label{table:3d-ohmi}      
\end{table}

We see in Table \ref{table:3d-ohmi} that resistivity does affect the number
of zonal flows produced in our simulations: two in O3N1 and one in O3N2,
instead of four in the $\eta_{\mathrm{O}}=0$ case of run B3L6. However, the outcome of runs O3N0 and O4N1 is the same as their 
non-resistive versions: a vortex in the former case, and one band in the latter. 

It is worth noting that from run O3N1 to O3N2, increasing the
resistivity by a factor ten resulted in an increase of Maxwell stress by a
factor two. Similarly, the total stress is twice higher in O4N3 than in the non-resistive equivalent scenario B4L6. 
This behavior comes from the diffusive broadening of the bands, leaving a wider region linearly unstable at their edges 
and maintaining a larger fraction of the domain unstable to the MRI. 

We thus find that for magnetic Lundquist numbers $\mathcal{S}_{\mathrm{O}} \gtrsim 1$, 
ohmic diffusion can reduce the number of zonal flows, but the system keeps self-organising. 
A similar final state is obtained when resuming the organised run B3L6 with resistivity: diffusion overcomes 
concentration between the bands and allows them to merge. The fact that a band remains at the end of the 
simulation suggests that this merging process is halted once the bands are too far to share magnetic flux. 
At this point, ohmic diffusion adds to turbulent diffusion in equation \eqref{eq:inucinde}, balancing the confinement by Maxwell stress.

\subsection{Ambipolar diffusion}

We finally look at the impact of ambipolar diffusion on self-organisation. With its non-linear dependence on $B$, this effect may lead to more complex behaviours than ohmic diffusion. The region of protoplanetary disks dynamically affected by ambipolar diffusion is expected to overlap the Hall dominated region, with typical ambipolar Elsasser numbers $\Lambda_{\mathrm{A}}$ below unity \citep{SLK15}. We first test the robustness of Hall self-organisation against ambipolar diffusion; we then turn our attention to a configuration in which \cite{BS14} report self-organisation without the Hall effect, and compare it to our Hall-dominated simulations. 

As previously, we activate ambipolar diffusion and the Hall effect simultaneously at $50 T_0$ and integrate until $200T_0$, with various values for the global magnetic flux, the ambipolar Lundquist number $\mathcal{S}_{\mathrm{A}}$ via the ratio $\eta_{\mathrm{A}} / v_A^2$, and the Hall parameter. The ambipolar Lundquist numbers are similar to the ohmic Lundquist numbers of the previous section, and the corresponding Elsasser numbers range from $0.01$ to $1$.

\subsubsection{Hall+Ambipolar MHD}

\begin{table}
\centering    
Simulations with ambipolar diffusion\\
\vspace{0.2cm}
\begin{tabular}{c c c c c c}        
\hline\hline                 
Name & $10^4 B_0$ & $\mathcal{S}_{\mathrm{A}} \:(\Lambda_{\mathrm{A}}^0) $ & $\mathcal{L}$ &
$\overline{\alpha}$ & state \\    
\hline                        
   A3N0 & $10$ & $25 \:(0.1)$ & $1$ & $1.0 \times 10^{-3}$ & 2 bands \\
   A3N1 & $10$ & $25 \:(0.1)$ & $2$ & $1.1 \times 10^{-3}$ & 3 bands \\	
   A3N2 & $10$ & $2.5 \:(0.01)$ & $1$ & $7.3 \times 10^{-4}$ & 2 bands \\	
   A3N3 &$10$ & $2.5 \:(0.01)$ & $0.4$ & $9.6 \times 10^{-4}$ & 2 bands \\   
   A3N4 &$10$ & $2.5 \:(0.01)$ & $0.2$ & $1.5 \times 10^{-3}$ & 2 bands \\   
   A4N5 &$1$ & $25 \:(0.01)$ & $1$ &  $1.3 \times 10^{-3}$ & channel \\
   A4N6 &$1$ & $25 \:(0.01)$ & $4$ &  $1.1 \times 10^{-3}$ & 3 bands \\   
   ABS	 &$18$ & $139\Omega \:(1)$ & $0$ & $4.2\times 10^{-3}$ & turbulent \\
   \hline                                   
\end{tabular}
\vspace{2mm}
\caption{Parameters for the 3D Hall+Ambipolar MHD runs: label of the run, mean
vertical magnetic field $B_0$, ambipolar Lundquist number $\mathcal{S}_{\mathrm{A}}$, corresponding ambipolar Elsasser number $\Lambda_{\mathrm{A}}^0$ at the inner radius $r_0=1$, Hall parameters $\mathcal{L}$, total turbulent stress $\overline{\alpha}$ and final state of the flow; ``channel'' describes a computational domain filled with linear HSI modes; only run ABS has a diffusivity scaling as $r^{3/2}$, corresponding to a constant ambipolar Elsasser number $\Lambda_{\mathrm{A}}=1$. }             %
\label{table:3d-ambi}      
\end{table}

We see in Table \ref{table:3d-ambi} that the total turbulent stress is not significantly affected by ambipolar diffusion in the low-transport state, the magnetic contribution still being negligibly small. Compared to the non-diffusive case, we find a smaller number of zonal flows with ambipolar diffusion in runs A3N0 to A3N2, but it is larger than the case of an equivalent ohmic Lundquist number (see Table \ref{table:3d-ohmi}). An inefficient merging of magnetic bands may be explained by the fact that ambipolar diffusion has little influence in the field-free region between the bands. 

When lowering the average magnetic flux in run A4N5, we obtain a disk filled with global channel modes and no zonal field. Since the average magnetic field is far below the threshold for HSI stabilisation, linear modes develop until the ambipolar diffusive damping rate, of order $\eta_{\mathrm{A}}/h^2$, balances their growth rate. The resulting stress has an almost flat radial profile, providing no confinement at the scale of our computational domain. Run A4N6 proves that when increasing $\mathcal{L}$ at the same field intensity, the Hall effect becomes competitive again and we recover several zonal fields. 

\begin{figure}[h]
\centering
\includegraphics[width=\hsize]{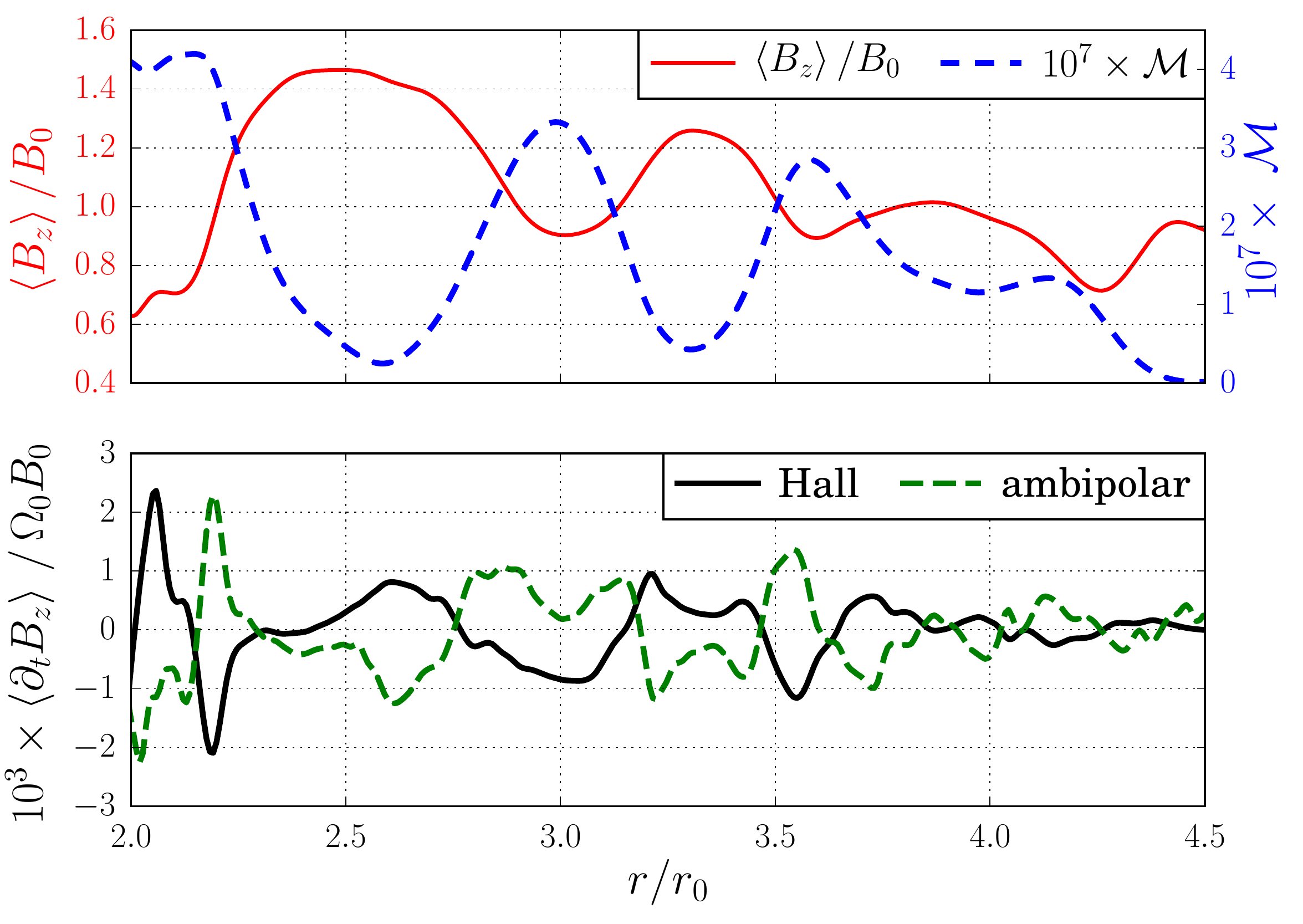}
\caption{Radial profiles in run A3N4, averaged in the vertical and azimuthal directions, and in time between $150T_0$ and $250T_0$. Upper panel: vertical magnetic field (solid red) and Maxwell stress (dashed blue); lower panel: contributions of the Hall (solid black) and ambipolar (dashed green) drifts to the induction of vertical magnetic field $\partial_t B_z$. }
\label{fig:snapambi}
\end{figure}

A new feature is the possibility of forming zonal fields at lower $\mathcal{L}<1$, as demonstrated by runs A3N3 and A3N4. 
We show in Fig.~\ref{fig:snapambi} the radial profile of vertical magnetic field in run A3N4, and the contributions of both non-ideal effects to the induction equation \eqref{eqn:dyn-b} in the vertical direction. The variations of $B_z$ have a much smaller amplitude than in Hall-only runs (see Fig.~\ref{fig:3d8bands}), but they still show a clear anti-correlation with the variations of Maxwell stress. As previously, we find that the Hall effect tries to increase $B_z$ in the bands and decrease it between. The ambipolar drift does precisely the opposite: it tries to smooth out the profile of $B_z$, and both contributions appear to cancel out on average. 

Since the Hall effect now balances ambipolar rather than turbulent diffusion, the self-organisation threshold given by \eqref{eqn:kl13crit} needs not hold anymore. Ambipolar diffusion acts to damp fluctuations in magnetic field, so the turbulent diffusivity $\eta_t$ entering equation \eqref{eq:inucinde} must eventually decrease when the ambipolar diffusivity is increased. In the limit of strong ambipolar diffusion, the bifurcation to a self-organised state should depend on $\eta_{\mathrm{A}}$ only. The fact that we observe two bands at $\mathcal{L}=0.2$ may result from an effective diffusivity $\eta_t + \eta_{\mathrm{A}}$ smaller than in the Hall-only case, due to the decrease of $\eta_t$ as a function of $\eta_{\mathrm{A}}$.

We conclude that self-organisation holds for ambipolar Elsasser numbers down to $\Lambda_{\mathrm{A}} = 0.01$, and that ambipolar diffusion can enable the formation of zonal flows for $\mathcal{L}$ as low as $0.2$.

\subsubsection{Self-organisation with ambipolar diffusion only}

Magnetic self-organisation was also reported by \cite{BS14} in local simulations of MRI
turbulent flows without the Hall effect; since ambipolar diffusion was
found to enhance magnetic flux concentration, we included this non-ideal
effect only and tried to reproduce the physical conditions of their run AD-4-64
within our global setup in run ABS. The initial magnetic field is set to $B_0 =
\sqrt{2 c^2 \rho_0 / 6400} \approx 1.8 \times 10^{-3}$, and the ambipolar Elsasser
number $\Lambda_{\mathrm{A}} = 1$ is constant throughout the computational domain, 
so that both setups should match at the inner boundary. Considering the narrowness of the magnetic structures observed
by \cite{BS14}, we used the HLLD approximate Riemann solver to reduce numerical
diffusivity in this run only. 

\begin{figure}[h]
\centering
\includegraphics[width=\hsize]{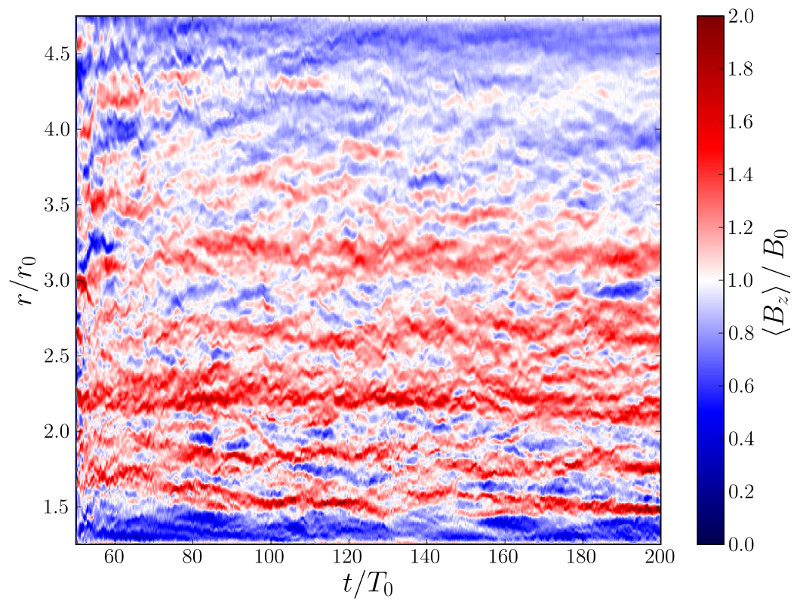}
\caption{Normalized vertical magnetic field $\brac{B_z} / B_0$ in run ABS. }
\label{fig:baiprof}
\end{figure}

In comparison to their figure 7, we confirm in our Fig.~\ref{fig:baiprof}
that the vertical magnetic flux does concentrate into bands. They are most
clearly visible in the inner half of the disk, where they live a few tens of orbits 
before being disrupted and formed again in the turbulent flow. Averaging this
profile in time, we find mean fluctuations of about $0.3 B_0$ revealing five bands at
radii $1.5$, $1.9$, $2.2$, $2.6$ and $3.2 r_0$. We note a broadening of these
bands with radius, suggesting some self-similar scaling $\lambda^2 \sim
\tilde{\eta}/ \Omega$ of the characteristic concentration length-scale $\lambda$ with the
local time-scale for some effective anti-diffusivity $\tilde{\eta}$, and not a
constant concentration length of order $h$ as for the Hall effect. 

\begin{figure}[h]
\centering
\includegraphics[width=\hsize]{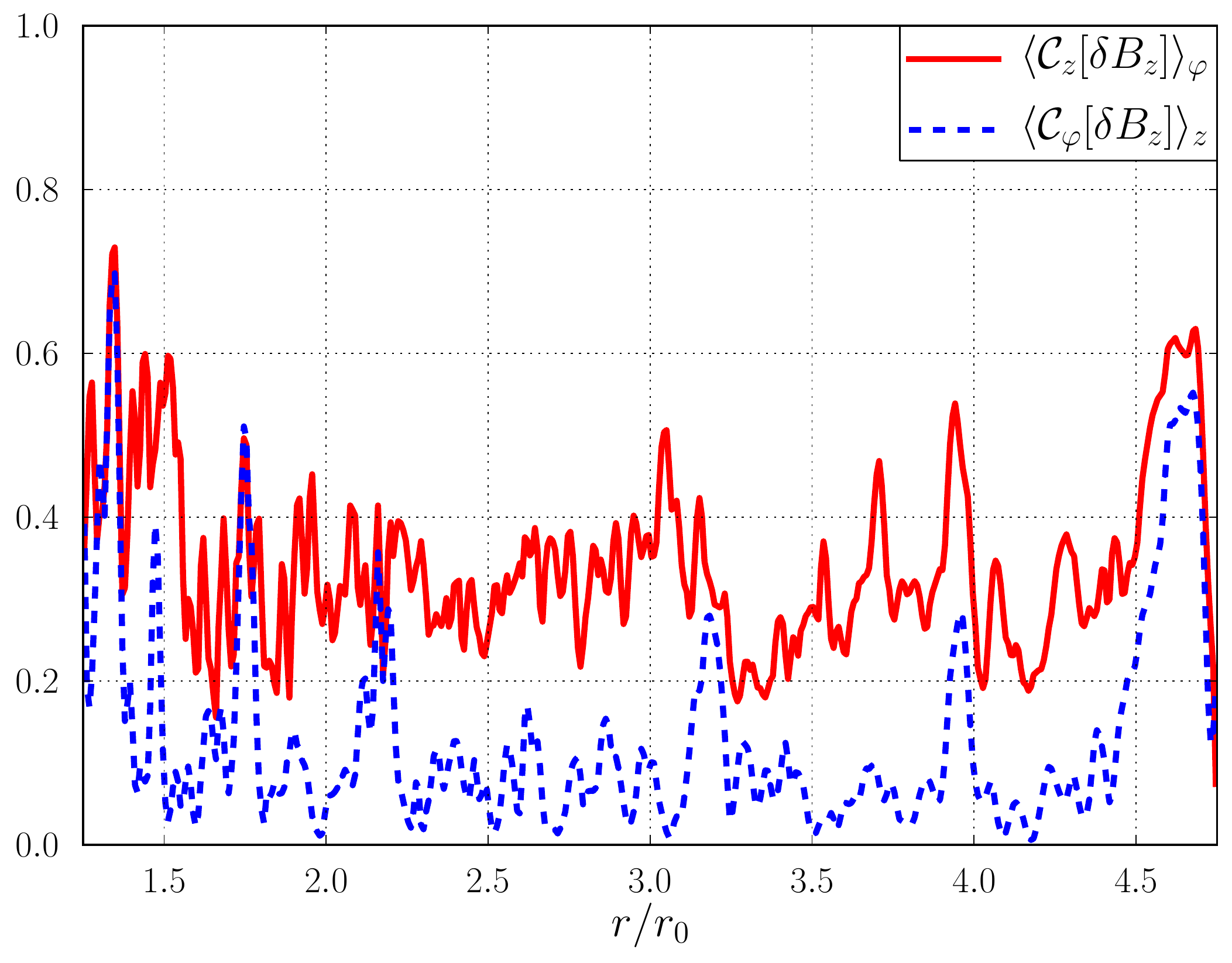}
\caption{Radial profiles of averaged vertical (solid red) and azimuthal (dashed
blue) correlations lengths for the fluctuation in vertical magnetic field
$\delta
B_z$ in run ABS at time $200T_0$. }
\label{fig:baicorr}
\end{figure}

We show in Fig.~\ref{fig:baicorr} the correlation profiles of $\delta B_z$ in
run ABS at time $200T_0$. The median values are $0.3$ and $0.08$ for the
vertical and azimuthal correlation lengths respectively, with peak values up to
$0.5$ in azimuthal correlation at the location of each spotted band. These
median values are consistent with those from a fully turbulent state (see
Fig.~\ref{fig:corryz}). In particular there is no strong vertical coherence in
the bands, confirming that these magnetic concentrations can never stabilize
the flow in the full height of the disk; in comparison, the zonal fields produced by the Hall effect
are always Hall-shear stable with vertical correlation factors close to unity. 

We conclude that the self-organisation process observed by \cite{BS14} is fundamentally
different from the Hall-induced self-organisation mechanism presented in this paper. 

\section{Summary and discussion}

We have presented a number of non-ideal MHD simulations dedicated to the Hall dominated regions of protoplanetary disks. We employed a global non-stratified model designed to naturally extend results from local shearing box simulations. For this purpose, we have extended the PLUTO code in order to include the Hall effect in cylindrical geometry, and validated our implementation in the linear regime of the resistive MRI. We then performed a series of global Hall-MHD simulations, focusing on the different stages of self-organisation found as the intensity of the Hall effect is increased. Finally, we addressed the consequences of a net toroidal magnetic flux, as well as ohmic or ambipolar diffusion, on this picture of Hall-induced self-organisation. 

	We can summarise our main findings as follow:
   \begin{enumerate}
      \item We confirm the transition from enhanced to lowered turbulent transport states when
increasing the intensity of the Hall effect; the turning point is found at $\mathcal{L}
\approx 0.1$, comparable to the estimate of KL13. 
      \item In the strong Hall limit $\mathcal{L} \gtrsim 1$, self-organisation leads to the formation of stationary zonal flows in cylindrical geometry; as the Hall length or the available magnetic flux is increased, so does the number of zonal flows, until the computational domain is filled with bands of accumulated magnetic field with typical width $h$. 
      \item When increasing $\mathcal{L}$ from zero, the transition from a small-scale turbulence to axisymmetric zonal flows is accompanied by the formation of large-scale magnetised vortices, a feature barely accessible in local simulations for it requires a large aspect ratio in the radial over vertical directions. 
	  \item Both the vortices and the zonal flows caused by the Hall effect may play a role in the process of planetary formation, since they can affect the rotation profile of the gas sufficiently to trap dust particles. 
      \item The self-organisation mechanism described in this paper holds when including a net toroidal field, ohmic or ambipolar diffusion; both diffusive effects cause the magnetic structures to merge over time, but do not change significantly the total torque exerted on the disk; ambipolar diffusion is found to enable self-organisation at lower Hall parameter $\mathcal{L}$. 
   \end{enumerate}

These results demonstrates that Hall-MRI is a very promising mechanism through which large scale structures can form and be observed. One should however keep in mind that these simulations are still very simplified and leave many questions opened. First, we have entirely neglected vertical stratification and we have simplified the radial structure to keep relatively constant diffusion parameters. Vertical stratification can be very important as it drives surface winds and accretion thanks to magnetic torques. Stratified shearing box models including the Hall effect have not shown any obvious self-organisation, but this is likely due to the limited radial extension of these boxes. The mechanism we propose for the self organisation is relatively simple and robust as it only relies on 3 hypotheses: 1- the disc is sufficiently ionised to be MRI/HSI unstable, 2- the $r-\phi$ component of the Maxwell stress vanishes for strong enough field strengths  and 3- the Hall length-scale is of the order of the disc scale height. Hypotheses 1 and 3 are usually satisfied in the midplane of stratified models. However, hypothesis 2 can be violated when a strong large scale wind is present, as the Maxwell stress is not only due to local physics but also to the global wind dynamics. Testing self-organisation in global stratified models is therefore essential to confirm this picture.

It is also worth pointing out that several of our simulations exhibit stable magnetised vortices. This might look surprising as vortices are known to be strongly unstable in magnetised discs due to magneto-elliptic instabilities \citep{MB09}. We have not investigated extensively the stability of these objects, but the presence of a strong field in the vortex core is most certainly the key to their stability, as this tends to rigidify flow motions and prevent fluid particles from entering into a resonance with the vortex turnover frequency.

Whether the structures produced in our simulations can be observed is yet another opened question. It is clear that some of these structures are dust traps, and should therefore accumulate millimetre-sized dust particles. However, this dust might also react on the flow, and even affect the ionisation structure, thereby changing the dynamics. This complex interplay between MHD, dust and chemistry have not been included in our models for obvious simplification reasons, but this step will be required if one is to predict the dust density contrast in rings and vortices and predict observational properties of these features.

\begin{acknowledgements}
We thank Mario Flock for helping us with the initial setup of our simulations and Matthew Kunz for useful discussions.
This  work  was  granted  access  to  the  HPC  resources  of IDRIS under  the allocation 2015-042231 made by GENCI. Some of the computations presented in this paper were performed using the Froggy platform of the CIMENT infrastructure (https://ciment.ujf-grenoble.fr), which is supported by the Rhône-Alpes region (GRANT CPER07\_13 CIRA), the OSUG@2020 labex (reference ANR10 LABX56) and the Equip@Meso project (reference ANR-10-EQPX-29-01) of the programme Investissements d'Avenir supervised by the Agence Nationale pour la Recherche.
\end{acknowledgements}


\bibliographystyle{aa}
\bibliography{hall_draft}

\begin{thebibliography}{45}
\expandafter\ifx\csname natexlab\endcsname\relax\def\natexlab#1{#1}\fi

\bibitem[{{ALMA Partnership} {et~al.}(2015){ALMA Partnership}, Brogan, Perez,
  Hunter, Dent, Hales, Hills, Corder, Fomalont, Vlahakis, Asaki, Barkats,
  Hirota, Hodge, Impellizzeri, Kneissl, Liuzzo, Lucas, Marcelino, Matsushita,
  Nakanishi, Phillips, Richards, Toledo, Aladro, Broguiere, Cortes, {Cortes, P.
  C.}, Espada, Galarza, Garcia-Appadoo, Guzman-Ramirez, Humphreys, Jung,
  Kameno, Laing, Leon, Marconi, Mignano, Nikolic, Nyman, Radiszcz, Remijan,
  Rodon, Sawada, Takahashi, Tilanus, Vila~Vilaro, Watson, Wiklind, Akiyama,
  Chapillon, de~Gregorio-Monsalvo, Di~Francesco, Gueth, Kawamura, Lee,
  Nguyen~Luong, Mangum, Pi{\'e}tu, Sanhueza, Saigo, Takakuwa, Ubach, van
  Kempen, Wootten, Castro-Carrizo, Francke, Gallardo, Garc{\'\i}a, Gonzalez,
  Hill, Kaminski, Kurono, Liu, Lopez, Morales, Plarre, Schieven, Testi, Videla,
  Villard, Andreani, Hibbard, \& Tatematsu}]{ALMA15}
{ALMA Partnership}, Brogan, C.~L., Perez, L.~M., {et~al.} 2015, \apjl, 808, L3

\bibitem[{Bai(2014)}]{B14}
Bai, X.-N. 2014, ApJ, 791, 137

\bibitem[{{Bai}(2015)}]{B15}
{Bai}, X.-N. 2015, \apj, 798, 84

\bibitem[{Bai \& Stone(2013)}]{BS13}
Bai, X.-N. \& Stone, J.~M. 2013, \apj, 769, 76

\bibitem[{{Bai} \& {Stone}(2014)}]{BS14}
{Bai}, X.-N. \& {Stone}, J.~M. 2014, \apj, 796, 31

\bibitem[{{Balbus} \& {Hawley}(1991)}]{BH91}
{Balbus}, S.~A. \& {Hawley}, J.~F. 1991, \apj, 376, 214

\bibitem[{{Balbus} \& {Papaloizou}(1999)}]{BP99}
{Balbus}, S.~A. \& {Papaloizou}, J.~C.~B. 1999, \apj, 521, 650

\bibitem[{{Balbus} \& {Terquem}(2001)}]{BT01}
{Balbus}, S.~A. \& {Terquem}, C. 2001, \apj, 552, 235

\bibitem[{{Barge} \& {Sommeria}(1995)}]{BS95}
{Barge}, P. \& {Sommeria}, J. 1995, \aap, 295, L1

\bibitem[{Benisty {et~al.}(2015)Benisty, Juh{\'a}sz, Boccaletti, Avenhaus,
  Milli, Thalmann, Dominik, Pinilla, Buenzli, Pohl, Beuzit, Birnstiel, de~Boer,
  Bonnefoy, Chauvin, Christiaens, Garufi, Grady, Henning, Huelamo, Isella,
  Langlois, M{\'e}nard, Mouillet, Olofsson, Pantin, Pinte, \& Pueyo}]{BJ15}
Benisty, M., Juh{\'a}sz, A., Boccaletti, A., {et~al.} 2015, \aap, 578, L6

\bibitem[{{Blaes} \& {Balbus}(1994)}]{BB94}
{Blaes}, O.~M. \& {Balbus}, S.~A. 1994, \apj, 421, 163

\bibitem[{{Chavanis}(2000)}]{C00}
{Chavanis}, P.~H. 2000, \aap, 356, 1089

\bibitem[{{Dipierro} {et~al.}(2015){Dipierro}, {Price}, {Laibe}, {Hirsh},
  {Cerioli}, \& {Lodato}}]{DP15}
{Dipierro}, G., {Price}, D., {Laibe}, G., {et~al.} 2015, \mnras, 453, L73

\bibitem[{{Evans} \& {Hawley}(1988)}]{EH88}
{Evans}, C.~R. \& {Hawley}, J.~F. 1988, \apj, 332, 659

\bibitem[{{Gammie}(1996)}]{G96}
{Gammie}, C.~F. 1996, \apj, 457, 355

\bibitem[{{Gressel} {et~al.}(2015){Gressel}, {Turner}, {Nelson}, \&
  {McNally}}]{GTNN15}
{Gressel}, O., {Turner}, N.~J., {Nelson}, R.~P., \& {McNally}, C.~P. 2015,
  \apj, 801, 84

\bibitem[{{Hawley} {et~al.}(1995){Hawley}, {Gammie}, \& {Balbus}}]{HGB95}
{Hawley}, J.~F., {Gammie}, C.~F., \& {Balbus}, S.~A. 1995, \apj, 440, 742

\bibitem[{Hayashi(1981)}]{H81}
Hayashi, C. 1981, Progress of Theoretical Physics Supplement, 70, 35

\bibitem[{{Jin}(1996)}]{J96}
{Jin}, L. 1996, \apj, 457, 798

\bibitem[{{Johansen} {et~al.}(2004){Johansen}, {Andersen}, \&
  {Brandenburg}}]{JAB04}
{Johansen}, A., {Andersen}, A.~C., \& {Brandenburg}, A. 2004, \aap, 417, 361

\bibitem[{{Kersal{\'e}} {et~al.}(2004){Kersal{\'e}}, {Hughes}, {Ogilvie},
  {Tobias}, \& {Weiss}}]{KH04}
{Kersal{\'e}}, E., {Hughes}, D.~W., {Ogilvie}, G.~I., {Tobias}, S.~M., \&
  {Weiss}, N.~O. 2004, \apj, 602, 892

\bibitem[{{Kunz}(2008)}]{K08}
{Kunz}, M.~W. 2008, \mnras, 385, 1494

\bibitem[{{Kunz} \& {Balbus}(2004)}]{KB04}
{Kunz}, M.~W. \& {Balbus}, S.~A. 2004, \mnras, 348, 355

\bibitem[{{Kunz} \& {Lesur}(2013)}]{KL13}
{Kunz}, M.~W. \& {Lesur}, G. 2013, \mnras, 434, 2295

\bibitem[{{Lesur} {et~al.}(2014){Lesur}, {Kunz}, \& {Fromang}}]{LKF14}
{Lesur}, G., {Kunz}, M.~W., \& {Fromang}, S. 2014, \aap, 566, A56

\bibitem[{{Lesur} \& {Longaretti}(2009)}]{LL09}
{Lesur}, G. \& {Longaretti}, P.-Y. 2009, \aap, 504, 309

\bibitem[{{Lovelace} {et~al.}(1999){Lovelace}, {Li}, {Colgate}, \&
  {Nelson}}]{LCN99}
{Lovelace}, R.~V.~E., {Li}, H., {Colgate}, S.~A., \& {Nelson}, A.~F. 1999,
  \apj, 513, 805

\bibitem[{{Mignone} {et~al.}(2007){Mignone}, {Bodo}, {Massaglia}, {Matsakos},
  {Tesileanu}, {Zanni}, \& {Ferrari}}]{M07}
{Mignone}, A., {Bodo}, G., {Massaglia}, S., {et~al.} 2007, in JENAM-2007, ''Our
  Non-Stable Universe'', 96--96

\bibitem[{{Mignone} {et~al.}(2012){Mignone}, {Flock}, {Stute}, {Kolb}, \&
  {Muscianisi}}]{MFSK12}
{Mignone}, A., {Flock}, M., {Stute}, M., {Kolb}, S.~M., \& {Muscianisi}, G.
  2012, \aap, 545, A152

\bibitem[{{Minoshima} {et~al.}(2015){Minoshima}, {Hirose}, \& {Sano}}]{MHS15}
{Minoshima}, T., {Hirose}, S., \& {Sano}, T. 2015, \apj, 808, 54

\bibitem[{{Mizerski} \& {Bajer}(2009)}]{MB09}
{Mizerski}, K.~A. \& {Bajer}, K. 2009, Journal of Fluid Mechanics, 632, 401

\bibitem[{Muto {et~al.}(2012)Muto, Grady, Hashimoto, Fukagawa, Hornbeck, Sitko,
  Russell, Werren, Cur{\'e}, Currie, Ohashi, Okamoto, Momose, Honda, Inutsuka,
  Takeuchi, Dong, Abe, Brandner, Brandt, Carson, Egner, Feldt, Fukue, Goto,
  Guyon, Hayano, Hayashi, Hayashi, Henning, Hodapp, Ishii, Iye, Janson,
  Kandori, Knapp, Kudo, Kusakabe, Kuzuhara, Matsuo, Mayama, McElwain, Miyama,
  Morino, Moro-Martin, Nishimura, Pyo, Serabyn, Suto, Suzuki, Takami, Takato,
  Terada, Thalmann, Tomono, Turner, Watanabe, Wisniewski, Yamada, Takami,
  Usuda, \& Tamura}]{M12}
Muto, T., Grady, C.~A., Hashimoto, J., {et~al.} 2012, The Astrophysical Journal
  Letters, 748, L22

\bibitem[{{O'Keefe}(2013)}]{K13}
{O'Keefe}, W. 2013, PhD thesis, Dublin City University

\bibitem[{{O'Keeffe} \& {Downes}(2014)}]{KD14}
{O'Keeffe}, W. \& {Downes}, T.~P. 2014, \mnras, 441, 571

\bibitem[{{Papaloizou} \& {Terquem}(1997)}]{PT97}
{Papaloizou}, J.~C.~B. \& {Terquem}, C. 1997, \mnras, 287, 771

\bibitem[{{Sano} \& {Stone}(2002)}]{SS02}
{Sano}, T. \& {Stone}, J.~M. 2002, \apj, 577, 534

\bibitem[{{Shakura} \& {Sunyaev}(1973)}]{SS73}
{Shakura}, N.~I. \& {Sunyaev}, R.~A. 1973, \aap, 24, 337

\bibitem[{Simon {et~al.}(2013)Simon, Bai, Armitage, Stone, \& Beckwith}]{SX13}
Simon, J.~B., Bai, X.-N., Armitage, P.~J., Stone, J.~M., \& Beckwith, K. 2013,
  ApJ, 775, 73

\bibitem[{Simon {et~al.}(2015)Simon, Lesur, Kunz, \& Armitage}]{SLK15}
Simon, J.~B., Lesur, G., Kunz, M.~W., \& Armitage, P.~J. 2015, MNRAS, 454, 1117

\bibitem[{{Tanga} {et~al.}(1996){Tanga}, {Babiano}, {Dubrulle}, \&
  {Provenzale}}]{TBDP96}
{Tanga}, P., {Babiano}, A., {Dubrulle}, B., \& {Provenzale}, A. 1996, \icarus,
  121, 158

\bibitem[{{Turner} {et~al.}(2014){Turner}, {Fromang}, {Gammie}, {Klahr},
  {Lesur}, {Wardle}, \& {Bai}}]{TFG14}
{Turner}, N.~J., {Fromang}, S., {Gammie}, C., {et~al.} 2014, Protostars and
  Planets VI, 411

\bibitem[{van~der Marel {et~al.}(2013)van~der Marel, van Dishoeck, Bruderer,
  Birnstiel, Pinilla, Dullemond, van Kempen, Schmalzl, Brown, Herczeg,
  Matthews, \& Geers}]{MD13}
van~der Marel, N., van Dishoeck, E.~F., Bruderer, S., {et~al.} 2013, Science,
  340, 1199

\bibitem[{{Wardle}(1999)}]{W99}
{Wardle}, M. 1999, \mnras, 307, 849

\bibitem[{{Wardle}(2007)}]{W07}
{Wardle}, M. 2007, \apss, 311, 35

\bibitem[{Weidenschilling(1977)}]{W77}
Weidenschilling, S.~J. 1977, \mnras, 180, 57

\end{thebibliography}

\begin{appendix}

\section{Spectral method for linear non-ideal MHD in axisymmetric configuration} \label{app:spectral}

We start with the full set of equations describing a magnetised fluid in
non-ideal compressible magnetohydrodynamics: 
\begin{align}
\partial_t \rho + \bm{\nabla \cdot} \left[ \rho \bm{v}\right] &= 0 \label{eqn:} \\
\rho \left(\partial_t \bm{v} + \bm{v \cdot\nabla v}\right) &= - \bnab P +
\bm{J \times B}  + \upsilon \bm{\Delta} \bm{v} - \rho\bnab\Phi \label{eqn:}
\\
\begin{split}
\partial_t \bm{B}&= \bm{\nabla \times} \left[ \bm{v \times B} - \eta \bm{\nabla\times B}
\right.\\
 &\left.- \lH \left( \bm{\nabla\times B} \right) \bm{\times B} + \gamma
\left(\left(\bm{\nabla\times B}\right) \bm{\times B} \right) \bm{\times B}
\right] \label{eqn:}
 \end{split}
\end{align}
completed by the closures $\bm{J} = \nabla \times \bm{B}$ and $P = c_s^2 \rho$
where $\rho$ is the fluid density, $\bm{v}$ is its velocity, $\bm{B}$ the
magnetic field, $P$ the pressure, $\bm{J}$ the electric current density, $\eta$
the ohmic diffusivity, $\upsilon$ the kinematic viscosity, $\lH$ the Hall
length and $\gamma$ the ambipolar diffusion coefficient. $\Phi$ is a
gravitational potential $\propto r^{-1}$ in cylindrical coordinates
$(r,\varphi,z)$. 

These equations are linearized about the stationary solution $\bar{\rho} =
\rho_0$, $\bar{B} = B_0 e_z$, $\bar{v} = \Omega r e_{\varphi}$ where $\Omega(r)
\propto r^q$ and $q=-3/2$ in the keplerian case. We restrict ourselves to
axisymmetric perturbations: $\partial_{\varphi} = 0$. Defining $D_r \equiv
\frac{1}{r} + \partial_r$, the equation for the perturbed density reads 
\begin{equation}
\partial_t \rho' = - \rho_0 \left( D_r  v'_r + \partial_z v'_z \right), 
\end{equation}
and the perturbed velocity: 
\begin{align}
\begin{split}
\partial_t v'_r &= - \frac{c_s^2}{\rho_0} \partial_r \rho' + \upsilon \left(
\partial_r^2 + \frac{1}{r}\partial_r \right) \left[ v'_r \right] \\
&+ 2\Omega v'_{\varphi} + \frac{B_0}{\rho_0}\left( \partial_z B'_r - \partial_r
B'_z \right) 
\end{split} \\
\partial_t v'_{\varphi} &= - \left(2+q\right) \Omega v'_r +
\frac{B_0}{\rho_0}\partial_z B'_{\varphi} \\
\partial_t v'_z &= -\frac{c_s^2}{\rho_0}\partial_z \rho' + \upsilon
\partial_z^2 v'_z
\end{align}
Finally for the perturbed magnetic field: 
\begin{align}
\begin{split}
\partial_t \bm{B'} &= \bm{\nabla \times} \left[ \Omega r \bm{e_{\varphi} \times
B'} - B_0 \bm{e_z \times v'} - \eta \bm{\nabla \times B'} \right. \\ 
&\left.+ \lH B_0 \bm{e_z \times} \left( \bm{\nabla \times B'} \right) + \gamma
B_0^2 \bm{e_z \times e_z \times} \left( \bm{\nabla \times B'}
\right)\right] 
\end{split}
\end{align}

\begin{figure}[h!]
\centering
\includegraphics[width=\hsize]{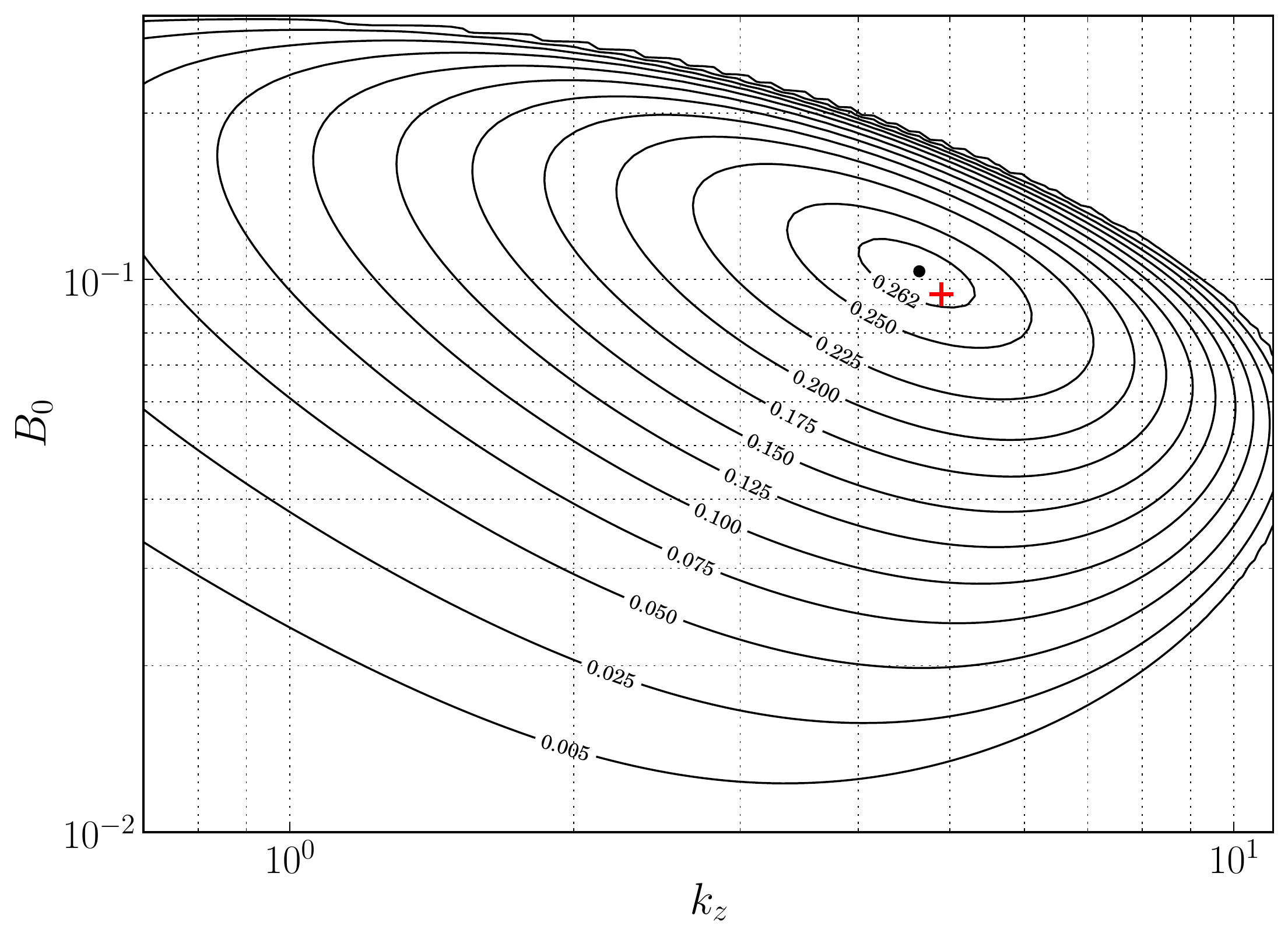}
\caption{Stability map with the parameters from \cite{KH04}. Our maximal growth
rate is marked with a black dot, theirs is marked with a red cross. }
\label{fig:stabkersale}
\end{figure}

Since we look for axisymmetric configurations, we can reduce the number of
variables and naturally ensure the $\nabla\cdot B=0$ condition using the vector
potential $A_{\varphi}$ such that $B_r = -\partial_z A_{\varphi}$ and $B_z =
D_r A_{\varphi}$. The linearized induction equations become: 
\begin{align}
\begin{split}
\partial_t B'_{\varphi} &= B_0 \partial_z v'_{\varphi} + \sigma \partial_z^2
B'_{\varphi} \\
&+ \lH B_0 \partial_z \left( \partial_z^2 - \partial_r D_r -
\partial_r\left[\Omega r \right] \right) \left[ A'_{\varphi} \right] \\
&+ \eta \left(\partial_r^2 + \frac{1}{r}\partial_r - \frac{1}{r^2} \right)
\left[ B'_{\varphi}\right]
\end{split}
\end{align}
\begin{equation}
\partial_t \left[-\partial_z A'_{\varphi}\right] = B_0 \partial_z v'_r -
\sigma\partial_z\left( \partial_z^2 + \partial_r D_r \right) \left[
A'_{\varphi}\right] + \lH B_0 \partial_z B'_{\varphi}
\end{equation}
with $\sigma \equiv \eta + \gamma B_0^2$. At this stage, we have a linear
system of partial differential equations for the the six fields $\bm{\xi}
\equiv (\rho', v'_r, v'_{\varphi}, v'_z, B'_{\varphi}, A'_{\varphi})$, which we
can write $\partial_t \,\bm{\xi} = \Delta \cdot \bm{\xi}$. Noticing that the
periodic $z$ coordinate appears only in derivatives, we can replace it
by a parameter $i k_z$ and reduce the problem to a system of one-dimensional
equations. Expanding over a finite set of Chebyshev polynomials, the extraction
of eigenmodes amounts to solving the generalized eigenvalue problem $\Delta
\cdot \bm{\xi_k} = \omega_k \Gamma \!\cdot \bm{\xi_k}$, where $\Gamma$ is
almost the identity matrix, in which a few lines are used to implement boundary
conditions. 

In order to test this method, we tried to reproduce the stability map from
figure 3.a of \cite{KH04}. The setup is the same as the one described in
\ref{sec:specmeth}, including viscosity and ohmic resistivity with
diffusivities $\nu = \eta = 0.003$ and no Hall effect. The maximal growth rate
is computed for different values of the background magnetic field $B_0$ and
vertical wave number $k_z$ of the eigenmode. The spectral resolution used for
this map is set to 32 modes after checking on several points that resolutions
of 256, 64 or even 16 spectral modes gave the same eigenvalue to better than
$10^{-4}$ relative accuracy. 

The resulting map is shown in Fig.~\ref{fig:stabkersale}. Our unstable domain
is geometrically very similar to the one of \cite{KH04}, but extends on a
larger portion of the $\left(k_z,B_0\right)$ plane and displays larger growth
rates. This comes from our different choice of boundary conditions for
$v_{\varphi}$ and $B_{\varphi}$. Our maximal growth rate $\gamma_M \approx
0.266 \Omega_0$ at $k_z = 4.64 r_0^{-1}$, $B_0=0.10$ is nevertheless very close
to the one they found in this plane.

\section{Numerical setup of \cite{KD14}\label{app:kd14}}
We present here the numerical setup of \cite{KD14} and make the connection of
their parameters real units to our dimensionless parameters.

\cite{KD14} use a cubic box of size $L_x=L_y=2.6$ and $L_z=0.195$. The setup is
a quarter disk with $r_\mathrm{int}=0.5$ and $r_\mathrm{ext}=2.58$. The units
are so that 1 numerical length unit corresponds to $5.2\,\mathrm{a.u.}$
\citep[][p 80]{K13}. This means, the disk extends from $r_0=2.6\,AU$ to
$r_1=13.4\,AU$.

This cylindrical disk is filled with a flow of uniform initial density with
$\rho=1.17\times 10^{-11}\,\mathrm{g.cm}^{-3}$ and a mean molecular mass
$m_\mathrm{n}=\,m_\mathrm{H}$  which gives a number density
$n=7\times10^{12}\,\mathrm{cm}^{-3}$. The ionisation fraction is assumed to be
$4\times 10^{-11}$ which implies a electron density of
$n_e=2.82\times10^2\,\mathrm{cm}^{-3}$. 

The initial field strength is chosen to be $50\,\mathrm{mG}$, so that the
Alfv\'en speed is 
\begin{align}
V_\mathrm{A}=\frac{B}{\sqrt{4\pi\rho}}=4.12\times 10^3\,\mathrm{cm.s}^{-1}.
\end{align}
The sound speed is not mentioned in \cite{KD14} but may be found in \cite{K13},
p.81: $c_s=8.04\times 10^4\,\mathrm{cm.s}^{-1}$. This sound speed is coherent
with a gas made of $\mathrm{H}_2$ at $T=130\,\mathrm{K}$.

With the values quoted above, we get the following dimensionless parameters
\begin{align}
\frac{h}{r_0}&=0.39,\\
\frac{c_s}{\Omega_0 r_0}&=4.35\times 10^{-2},\\
\frac{V_\mathrm{A}}{\Omega_0 r_0}&=2.23\times 10^{-3},\\
\frac{\ell_H}{r_0}\equiv \frac{c\sqrt{\rho \pi}}{2\pi e n_e h}&=5.5\times
10^{-3}.
\end{align}
The surprising low relative strength of the Hall effect ($\ell_H/h=1.4\times
10^{-2}\ll 1$) comes from the high ionisation fraction and the low midplane
density compared to other models \citep[e.g.][]{W07}, and the large geometrical
thickness compared to the pressure scale height ($h\Omega_0/c_s=9$)

\end{appendix}

\end{document}